\documentclass[fleqn,usenatbib]{mnras}

\usepackage[T1]{fontenc}

\DeclareRobustCommand{\VAN}[3]{#2}
\let\VANthebibliography\thebibliography
\def\thebibliography{\DeclareRobustCommand{\VAN}[3]{##3}\VANthebibliography}

\usepackage{booktabs} 
\usepackage{float} 
\restylefloat{table} 
\usepackage{multicol} 
\usepackage{dblfloatfix} 

\usepackage{graphicx}	
\usepackage{amsmath}	
\usepackage{amssymb}	

\usepackage[dvipsnames]{xcolor}

\usepackage{caption}

\usepackage{newtxtext,newtxmath}

\title[Curvature in 6dFGS, BOSS and eBOSS]{Full-Shape Galaxy Power Spectra and the Curvature Tension}

\author[]{
Aaron Glanville $^{1}$\thanks{E-mail: a.glanville@uq.edu.au} , Cullan Howlett $^{1}$, Tamara Davis $^{1}$
\\
$^{1}$The University of Queensland, School of Mathematics and Physics, QLD 4072, Australia
}

\date{Accepted XXX. Received YYY; in original form ZZZ}

\pubyear{2022}

\begin{document}
\label{firstpage}
\pagerange{\pageref{firstpage}--\pageref{lastpage}}
\maketitle

\begin{abstract}
With recent evidence for a possible ``curvature tension" among early and late universe cosmological probes, Effective Field Theories of Large Scale Structure (EFTofLSS) have emerged as a promising new framework to generate constraints on $\Omega_k$ that are independent of both CMB measurements, and some of the assumptions of flatness that enter into other large-scale structure analyses. In this work we use EFTofLSS to simultaneously constrain measurements from the 6dFGS, BOSS, and eBOSS catalogues, representing the most expansive full-shape investigation of curvature to date. Fitting the full-shape data with a BBN prior on $\Omega_b h^2$ and fixed $n_s$, we measure $\Omega_k = -0.089^{+0.049}_{-0.046}$, corresponding to a $\sim 2 \sigma$ preference for curvature. We argue that this result cannot be biased towards flatness by assumptions in the fitting methodology. Using the Bayesian evidence ratio our full-shape data assigns betting odds of 2:1 in favour of curvature, indicating present measurements remain broadly compatible with both flat and curved cosmological models. When our full-shape sample is combined with Planck 2018 CMB measurements, we break the geometric degeneracy and recover a joint fit on $\Omega_k$ of $-0.0041^{+0.0026}_{-0.0021}$. Using the suspiciousness statistic (built on the standard Bayes factor), we find evidence for a moderate tension between Planck 2018 and our suite of full-shape measurements, at a significance of $1.76 ^{+0.14}_{-0.11} \sigma$ ($p \sim 0.08 \pm 0.02$). These results demonstrate the usefulness of full-shape clustering measurements as a CMB independent probe of curvature in the ongoing curvature tension debate.
\end{abstract}




\section{Introduction}

Over the past decade, remarkable improvements in the precision and accuracy of our constraints have exposed clear inconsistencies between cosmological datasets. While much discussion has centred around the consistency of early and late universe measurements in their constraints of $H_0$, recent attention has been drawn to a potential tension among early universe curvature constraints \citep{Ooba_2019, Handley2019, Park_2019, DiValentino_2020, Efstathiou2020, Vagnozzi2020}. Determining what is driving these inconsistencies, and how to ameliorate the tensions which arise, remains a pressing challenge in modern cosmology. When taken alone, constraints made using the fiducial Planck measurements and likelihoods exhibit a strong preference for closed universes, in contrast to other early universe probes of curvature \citep{Planck_2018, DiValentino_2020}. This apparent discrepancy has raised questions regarding the internal consistency of data sets employed in joint constraints that use CMB data, and the confidence which can be given to such joint fits \citep{Handley2019, DiValentino_2020, Efstathiou2020}. With recent advances in modelling the full shape of the matter power spectrum, a powerful new framework has emerged to generate competitve curvature constraints directly from clustering measurements, independently of CMB information \citep{Chudaykin_2021}. 

One of the most promising of these full-shape models is the state-of-the art effective field theory of large scale structure (hereafter EFTofLSS), which reliably models the power spectrum of biased matter-density tracers up to mildly non-linear regimes ($k \sim 0.2$) \citep{dAmico_2020, Chudaykin_2020}. This is in contrast to methods which fit the imprinted BAO feature exclusively \citep{Beutler_2011, Beutler_2017, eboss_2020}, or fit the full-shape of the power spectrum using template-based methods (where the template is usually based on a fiducial flat-$\Lambda$CDM cosmology, \citealt{Tegmark2006, Sanchez2008, Reid2010, Parkinson2012}). While such mildly non-linear models provide a relatively modest extension in k-space (from $k_{\rm{max}} \sim 0.1$ to $k_{\rm{max}} \sim 0.2$), the cubic scaling of $n_{\rm{modes}}$ with $k_{\rm{max}}$ opens up a substantial amount of information for cosmological model fitting. 

In this work, we combine power spectrum measurements from the 6dFGS, BOSS, and eBOSS surveys in the most comprehensive full-shape analysis of curvature to date \citep{6dF_paper, Alam_2017, eboss_2020}\footnote{While drafting this manuscript, \cite{Neveux_2022} also released a full-shape analysis of the combined BOSS and eBOSS catalogues, although their work focuses on flat cosmological models and uses a different theoretical model and fitting methodology.}. We use this combined sample to constrain both flat and curved $\Lambda$CDM cosmological models, with minimal external assumptions or priors. Using the Bayesian evidence ratio, we quantify the statistical preference for curved/flat cosmological models exhibited by these full-shape measurements in isolation. By strictly analysing the full-shape data alone, we are able to disentangle the cosmological information provided by clustering measurements in isolation, allowing us to understand the statistical confidence of joint fits with external probes. We then determine the consistency of these results with Planck CMB measurements, using the suspiciousness statistic \citep{Handley_Lemos2019, Handley2019}.


In order to meaningfully explore the significance of this possible tension, it is also critical to quantify any aspects of our analysis that could potentially bias our results towards flatness. In particular, the support for flatness (within $\pm 10\%$) from early CMB experiments, along with the success of the inflationary model (which predicts flatness at late times, \citealt{Guth_81, Linde_82, Jaffe_2001, Netterfield_2002}) have made the assumption of flatness a convenient axiom in large scale structure physics. Many cosmological analysis tools and techniques also assume flatness as a matter of theoretical and computational simplicity. As such, it is imperative that we guard against confirmation bias by rigorously investigating the impact of these common assumptions in all steps of our analysis, from observation to results. In this work, we systematically review the potential assumptions of flatness which can arise in measurements of large scale structure, and determine the biases these can introduce in the emerging curvature constraints provided by EFTofLSS. 

Our paper is structured as follows. Through Section 2, we expand on the current state of the curvature tension,  provide some background for the EFTofLSS models used in this work, and justify our choice to fit the full-shape directly by detailing some common points where flatness is assumed in normal measurements of large scale structure. In Section 3, we detail the methodology behind parameter constraints and model comparisons used in this research, alongside the datasets employed in our work. In Section 4, we validate the accuracy of our full-shape pipeline (and in particular, fits to $\Omega_k$) using the publicly available Nseries mock catalogues. We then examine the common assumption of flatness in the choice of fiducial cosmology, to identify if this could introduce any substantial bias in fits to full-shape clustering measurements. In Section 5 we use our pipeline to simultaneously fit clustering measurements from the 6dFGS, BOSS, and eBOSS catalogues providing the most robust full-shape only constraints of $\Omega_k$ to date. Using the Bayesian evidence ratio, we then measure the statistical preference for flatness/curvature exhibited by our suite of full-shape measurements in isolation. Finally, we combine full-shape measurements with Planck 2018 CMB information, first recovering their joint posterior, and then determining the consistency of these datasets using the suspiciousness statistic.

\section{Background}

\subsection{Curvature Tension}

In conjunction with the axioms that form the standard $\Lambda$CDM model, spatial flatness has been a common assumption in cosmological analysis for some time. This preference is supported by the overwhelming body of evidence (across early and late universe probes), which suggest the universe is either spatially flat, or remarkably close to flat. Planck temperature and polarization spectra, when combined with lensing reconstruction information, yield fits which are  consistent with flatness to within $\sim 1\%$ ($\Omega_k = -0.0106 \pm 0.0065$). These constraints are further sharpened with the addition of BAO measurements, yielding $\Omega_k = 0.0007 \pm 0.0019$, corresponding to a $1 \sigma$ detection of flatness at an accuracy of $0.2\%$ \citep{Alam_2017, Planck_2018}. This preference for flatness is also supported amongst joint fits with other late-universe cosmological probes --- constraints using CMB spectra in combination with type Ia supernovae from Panetheon once again recover flatness to within $1 \sigma$ ($\Omega_k = -0.0061 ^{+ 0.0062}_{-0.0054}$). 

While fits which combine CMB measurements with external probes exhibit a consistent preference for flatness, CMB measurements in isolation exhibit a stark preference for closed cosmologies (measured as $\Omega_k = -0.044^{+0.018}_{-0.015}$ in \citealt{Planck_2018}). This is broadly driven by a preference for models with a higher than expected amount of lensing, with the most recent measurements recovering a lensing amplitude which exceeds theoretical forecasts by $\sim 2 \sigma$. In the absence of any external information (most notably lensing reconstruction), closed cosmologies provide a natural explanation for this lensing excess via an increase in $\Omega_m$. This preference for high lensing amplitudes and closed cosmologies within CMB measurements has been extensively discussed in the literature, with some evidence to suggest this effect may simply be a statistical fluctuation \citep{Planck_2015, Planck_2018, Efstathiou_2021}. In revisiting curvature constraints from Planck information, \cite{Efstathiou_2019} show the inclusion of $\Omega_k$ primarily modulates the $\chi^2$ values from the $\mathcal{\ell} < 30$ \textsc{Commander} temperature likelihood. When removing the $\mathcal{\ell} < 30$ multipoles, \cite{Efstathiou_2019} argue models with a free $\Omega_k$ do not perform substantially better than flat models, as measured by their respective $\chi^2$ values. Additionally, Planck measurements exhibit a strong preference for inflationary models with a large number of e-foldings, inconsistent with models of incomplete inflation which may naturally result in closed cosmologies.

Several recent papers have argued that the significance of this preference for closed universes, alongside the ubiquity of joint likelihoods which include CMB information, warrant further investigation.  \cite{DiValentino_2020} provide evidence for a substantial discord between CMB measurements and a host of late universe probes (including BAO, Supernovae, lensing, and cosmic shear measurements) in $\Lambda$CDM models which permit curvature. This is supported by \cite{Handley2019}, who determine Planck data is inconsistent with BAO and lensing measurements at a significance of $\leq 2.5 \sigma$, and consistent with a closed universe (assigning Bayesian betting odds of 50:1 against flatness). This comparison was further expanded in \cite{Vagnozzi2020}, who identify evidence of a substantial tension between full-shape clustering measurements (evaluated up to $k_{\textrm{max}} = 0.135$), and CMB spectra in models which permit curvature. While the origin of this discord continues to be discussed, any major inconsistency between the Planck measurements and our suite of late-universe probes represent a substantial challenge for our standard approach to fitting $\Omega_k$. Precise measurements of $\Omega_k$ (as part of the standard $\Lambda$CDM model) almost always rely on information from a combination of cosmic probes. As the earliest available measurement of cosmic structure, CMB measurements provide an invaluable ``lever-arm'' across cosmic history to break parameter degeneracies in late-universe measurements, making CMB measurements a ubiquitous element of curvature constraints. Implicit in the combination of multiple datasets is the assumption that all measurements contained within the joint likelihood could plausibly arise from the same universe. Evidence of a significant incompatibility within the CMB data, regardless of its origin, opens serious questions on the reliability of these joint fits. 

As the potential significance of these apparent inconsistencies continues to be discussed, it is more important than ever to generate a range of robust curvature constraints that do not rely on CMB measurements. Measurements of the BAO feature provide one such independent probe of cosmic curvature, with the recent eBOSS survey recovering flatness to within $1 \sigma$ \citep[$\Omega_k = 0.078^{+0.086}_{-0.099};$][]{eboss_2020}. Additionally, advances in fitting the broadband power spectrum up to non-linear scales have offered a valuable new perspective in this discussion, which has already been capitalised on in recent work. Most notably, \citet{Chudaykin_2021} show that combining post-reconstruction BAO information with full-shape measurements allows for competitive constraints on curvature ($\Omega_k = -0.043 \pm 0.036$), which are wholly independent of CMB information. 

\subsection{Flatness Assumptions in Large Scale Structure}

With the emergence of the first reliable CMB anisotropy measurements two decades ago, a flat universe (within $\pm 10\%$) has been consistently preferred \citep{Balbi_2000, Jaffe_2001, Masi_2002}. The significant reduction in computational time afforded by fixing $\Omega_k = 0$, in conjunction with the strong body of observational and theoretical evidence supporting a flat universe, has made flatness a very common ``short-cut'' assumption in cosmological analyses. The ubiquity of even implicit assumptions of flatness (as a default in many cosmological analysis tools or data products), makes it important to guard against confirmation bias when considering curvature constraints provided by new models such as EFTofLSS. As such, we outline some of the more significant assumptions of flatness in large scale structure measurements and analyses, in order to systematically evaluate whether these introduce any significant bias in our EFTofLSS curvature constraints.

\subsubsection{Fiducial Cosmology}
\label{sec:fiducial_cosmology}

Measurements of cosmic structure begin with a galaxy redshift catalogue, which details the angular sky position and redshift of hundreds of thousands of individual galaxies. In order to measure clustering statistics from this sample, a fiducial cosmology must be employed to convert observed redshift to comoving distances. While measurements of cosmic structure (such as the BAO feature) are broadly robust to systematics in redshift measurements themselves \citep{Glanville_2021}, offsets in fiducial cosmology ($\sim 0.5\%$ in $\Omega_{\rm m}$) from the underlying sample truth are known to introduce biases in anisotropic BAO fits of $\sim 0.1\%$ \citep{VargasMagana_2018}. For most realistic settings this effect is firmly sub-dominant to statistical errors, and does not substantially affect parameter constraints when propagated through BAO analysis \citep{GilMarin2020}. Model independent approaches to fitting the BAO feature have been proposed in the literature, most notably by measuring the radial two-point correlation function \citep{Sanchez_2010, Sanchez_2013}, however such techniques come at the expense of a dampened BAO signal. While the increased scale of next generation surveys such as the Dark Energy Spectroscopic Instrument \citep[DESI;][]{DESI2016} may allow for competitive model independent fits of the BAO feature, at present most major investigations rely on fiducial cosmologies which are overwhelmingly flat \citep{Beutler_2011, Parkinson2012, Alam_2017, Bautista_2021, GilMarin2020, Tamone_2020, DeMattia_2021, Hou_2021, Neveux_2020}. In this work, we systematically investigate the potential biases introduced by this assumption by ``stress-testing'' our fits --- that is, by constraining clustering measurements made using non-flat fiducial cosmologies, for comparison with a flat baseline measurement.

\subsubsection{Reconstruction}
\label{sec:Reconstruction}
In BAO analyses, it is common practice to use reconstruction techniques to improve the clarity of the BAO feature. Broadly, the non-linear evolution of galaxies arising from gravitational effects work to dampen the amplitude of the BAO wiggles, weakening the observed signal. Reconstruction reduces this signal loss by using the observed density field to infer the local velocity field, partially reversing these non-linear motions to generate a ``reconstructed'' sample. Importantly, reconstruction techniques must assume a fiducial cosmology (when converting from observed redshifts to comoving distances), a fiducial bias parameter in the modelling of the underlying density field from the observed galaxy distribution, and a fiducial growth rate of structure to account for redshift-space distortions.

Reconstruction significantly improves the precision of BAO distance measurements, and the fiducial cosmology of reconstruction only has a small impact at this scale  \citep{Padmanabhan_2012, Burden_2014, Kazin_2014, Carter_2020}. However, there is evidence to suggest that the use of a fiducial reconstruction cosmology which deviates from the underlying cosmology of a sample can bias the monopole shape and quadrupole amplitude, at the $1\%$ and $5\%$ levels respectively \citep{Sherwin_2019}. As such, the use of reconstruction on observed data (where the underlying sample cosmology is not known) represents a source of potential bias for fits to the full shape of the power spectrum. 

Additionally, it is common practice to fold in post-reconstruction BAO information with other large scale structure measurements to improve joint constraints. In the context of EFTofLSS, this approach was taken by \cite{Chudaykin_2021}, who used post-reconstruction BAO information in combination with full-shape data to constrain a range of extensions to the standard model (such as oCDM, $w_0 {\rm CDM}$ and $w_0 w_a {\rm CDM}$ models). In these fits, \cite{Chudaykin_2021} correctly note this post-reconstruction BAO information is covariant with the pre-reconstruction power spectrum. However, it is important to first verify that pre-reconstruction full-shape and post-reconstruction BAO measurement are internally consistent, before we can confidently combine this information. At the cost of a reduction in our constraining power, we therefore opt to disentangle the assumptions of reconstruction from our analysis by fitting full-shape measurements independently of post-reconstruction BAO information. This means that there are no assumptions of flatness entering in our analysis due to reconstruction.

\subsubsection{Covariance Matrix}
\label{sec:mock_catalogues}
When estimating the errors associated with binned clustering measurements, one must compute the covariance matrix for this sample from a large number of mock catalogues. Due to the high computational overhead associated with generating these mocks, both the underlying cosmology of these catalogues, and the fiducial cosmology used to compute comoving distances, are generally fixed to $\Omega_k = 0$ \citep{Kitaura_2016, RodriguezTorres_2016, RodriguezTorres2017, Favole_2016, Favole_2017}.\footnote{The recently completed Quijote N-body simulations provide a suite of 11,000 mock power spectra generated from universes which extend over a hypercube of parameters, including $\Omega_{k}$ \cite{VillaescusaNavarro_2020}.} 

Even at linear order, the covariance matrix is proportional to the power spectrum and shot-noise, and hence is dependent on cosmology. However, in practice the clustering from simulations is tuned to match the amplitude of the measured power spectrum (and often bispectrum) of the data because the galaxy bias is unknown. As we demonstrate  in Section \ref{Sec:Results} and Figure~\ref{fig:Nseries_varying_fiducial}, changing $\Omega_{k}$ mainly also changes the amplitude of the power spectrum (and is very degenerate with the galaxy bias). Hence, we expect the covariance matrix to be only very weakly dependent on the choice of $\Omega_{k}$ used to produce the simulations. 

To expound further, consider if the true Universe had, e.g., $\Omega_{k}=0.1$. Imagine we then generated two sets of simulations and two covariance matrices from those simulations, one with $\Omega_{k}=0.1$ and the other assuming flatness. These would both get tuned to match the clustering in the data, with the result that the simulations with $\Omega_{k}=0.1$ would be tuned to have a lower value for the galaxy bias. However, in both cases, the resulting covariance matrices would be very similar. 

We hence argue that the choice of cosmology used to generated the covariance matrix is very unlikely to bias measurements from large-scale structure towards flatness.

\subsubsection{Summary}

Galaxy clustering measurements, and the statistics derived from these, assume spatial flatness at all stages of analysis --- at the catalogue level (with the use of reconstruction algorithms),  through our analysis pipelines (by assuming flat fiducial cosmologies), and in our final model fitting (through assumptions in common cosmological codes, and model choices in computing the covariance matrix). In order to produce results as independent of these assumptions as possible, we purposefully choose to fit only pre-reconstruction full-shape data, we fit directly for both flat and non-flat cosmological models, and we use mock catalogues to quantify the impact of assuming a consistent non-flat fiducial cosmology in our analysis pipeline.

\subsection{Effective Field Theory of Large Scale Structure}

\subsubsection{Theoretical Overview}

EFTofLSS comes from a rich theoretical background, initially developed to model the back reactions from small scale inhomogeneities, to determine their impact on spacetime at larger scales \citep{Baumann_2012}. As improvements in observation have pushed cosmology to a precision science, the EFTofLSS model has evolved into a framework that can provide highly accurate theoretical predictions of the power spectrum, which are directly comparable to the clustering statistics derived from biased tracers of the matter field \citep{Carrasco_2012, Porto_2014, Senatore_2015}. While EFTofLSS initially only accommodated models of the dark matter density field, the equations of motion associated with the baryon, massive neutrino, and dark energy density fields have since been incorporated into a unified formalism \citep{Lewandowski_2015, Senatore2017effective, Lewandowski_2017}. We use this section to provide a brief theoretical overview of the EFTofLSS model used in this work, in order to motivate some model choices in our analysis. For a more detailed description of the EFTofLSS model and its theoretical derivation, we direct the reader to \cite{dAmico_2020} and \cite{Chudaykin_2020}.

Broadly, EFTofLSS models use physically motivated perturbations of the linear power spectrum to evaluate the power spectrum in the non-linear regimes. To derive these non-linear perturbations, each component of the cosmic density (e.g. dark matter, baryonic matter, neutrinos etc.) is modelled as a distinct, fluid-like field. The clustering contributions from each component field may be computed using perturbative expansions in powers of $k/k_{\rm{nl}}$, where $k$ is the mode of interest, and $k_{\rm{nl}}$ is a non-linear cut-off scale. The contribution of each perturbation to the overall field is weighted by a series of bias parameters, which parameterise effects such as the linear galaxy bias, shot-noise, and counter-terms which encode the effects of back-reactions on large-scales from modes below the cut-off scale which have otherwise been ignored. In order to avoid modelling complex and computationally expensive small scale non-linear effects, EFTofLSS models \citep[such as][]{dAmico_2020, Ivanov_2020} frequently restrict themselves to the linear ($k < 0.1$) and mildly non-linear ($ 0.1 < k < 0.2$) regimes. Despite the relatively small extension in $k$ space associated with fitting this mildly non-linear regime, the cubic scaling of $n_{\rm{modes}}$ with $k_{\rm{max}}$ ensures even small improvements in scale resolution open up a large number of new density modes for analysis. Since most galaxy clustering statistics already provide measurements at these mildly non-linear scales, EFTofLSS models can massively extend the useful information that can be extracted from these previously unconstrained modes. 

A number of codes now exist that implement EFTofLSS theories to model clustering statistics from an ensemble cosmological model, most notably the \textsc{PyBird} and \textsc{CLASS-PT} codes \citep{dAmico_2020, Chudaykin_2020}. These codes have already been used to provide competitive, independent constraints on a range of cosmological parameters directly from the BOSS DR12 power spectra, with blinded challenges verifying the robustness of these results \citep{dAmico_2020, Philcox_2020, Nishimichi_2020}. EFT models have also been used to place constraints on a number of common extensions to the standard, flat $\Lambda$CDM model. Notably, \cite{Chudaykin_2021} used full-shape measurements in conjunction with post-reconstruction BAO and supernova information to constrain curved $\Lambda$CDM models, yielding results consistent with flatness at just over $1~\sigma$ ($\Omega_k = -0.043 \pm 0.036$). A number of investigations have also used full-shape power spectrum measurements in conjunction with the EFTofLSS model to measure the dark energy equation of state parameter, and the sum of neutrino masses \citep{Chudaykin_2020, Ivanov_2020_Neutrino, dAmico_2021, Zhang_2021_Pybird_neutrino}.

\subsubsection{Model Parameterisation and Marginalisation}
\label{sec:Model Parameterisation}
We use this section to provide a brief overview of the technical choices made in this analysis which relate to our full-shape model. In this work we use the EFTofLSS model incorporated in the \textsc{PyBird} code, which models power spectra up to 1-loop order (i.e. up to  $3^{\rm{rd}}$ order in perturbations), introducing a total of 10 EFT bias parameters for each sample in our analysis. Four of these terms correspond to the galaxy bias parameters $b_{\rm{i}}$, which arise from the expansion of the galaxy density and velocity fields in terms of their dark matter counterparts. The small scale (or UV) dependence of redshift space distortions is cancelled through counter-terms parameterised as $c_{\rm{r, 1}}$ and $c_{\rm{r, 2}}$, and non-linearities in the dark matter density are similarly accounted for by the $c_{\mathrm{ct}}$ counter-term, which models the effective sound speed of dark matter. This leaves 3 remaining stochastic bias parameters, $c_{\epsilon, \rm{i}}$, which are shot-noise terms that absorb the difference between the \textit{observed} noisy galaxy distribution and the expectation values of the underlying fields. 

Following the procedure of \cite{dAmico_2020}, these bias parameters can be separated into terms which appear at linear order in the power spectrum, and higher-order terms. The 7 counter-terms which appear at linear order may be analytically marginalised when computing our likelihood,  leaving \{$b_1, b_2, b_4$\} as free bias parameters. \cite{dAmico_2020} find that $b_2$ and $b_4$ are almost completely anti-correlated in simulations and data. As such, they perform a change of variables from \{$b_2$, $b_4$\} to \{$c_2$, $c_4$\}, where:
    \begin{equation}
        c_2 = \dfrac{1}{\sqrt{2}} (b_2 + b_4)
    \end{equation}
    \begin{equation}
        c_4 = \dfrac{1}{\sqrt{2}} (b_2 - b_4)
    \end{equation}

In this work we follow this convention, performing a change of variables from \{$b_2$, $b_4$\} to \{$c_2$, $c_4$\}, and marginalising over the 7 bias terms which appear at linear order. \cite{dAmico_2020} determine that for BOSS-like volumes, the functions that are scaled by $c_4$ are small enough to be negligible, and as such they fix $c_4$ to 0 in their work. We do the same, and as such, our standard fitting procedure consists of our chosen suite of cosmological parameters (e.g \{$\ln(10^{10} A_{s}), \ h, \ \Omega_{\textit{cdm}} h^2, \ \Omega_{b} h^2, \ n_s, \ \Omega_{\rm k}$\}), with 2 additional bias parameters \{$b_1$, $c_2$\} for each independent clustering sample. 

\section{Methodology}

\begin{table*}[h]
\centering
\renewcommand{\arraystretch}{1.2} 
\setlength{\tabcolsep}{3.0pt}
\begin{tabular}{@{}llllll@{}}
\toprule
 & $\{z_{\rm{min}} - z_{\rm{max}}\}$ & $z_{\rm{eff}}$ & $N_{\rm{sources}}$ & $V_{\rm{eff}} \ (\rm{Gpc}^3)$ & $k_{\rm{max}} \ \{ k_0, k_2, k_4 \}$ \\ \midrule
 6dFGS DR3 & $\{0.01 - 0.2\}$ & $0.096$ & $75 \ 117$ & $0.12$ & $\{0.15, \ 0.15, \ 0.1 \}$ \\ 
 BOSS z1 NGC & $\{0.2 - 0.5\}$ & $0.38$ & $429 \ 182$ & $2.6$ & $\{0.2, \ 0.2, \ 0.15 \}$ \\ 
 BOSS z1 SGC & $\{0.2 - 0.5\}$ & $0.38$ & $174 \ 819$ & $1.0$ & $\{0.2, \ 0.2, \ 0.15 \}$ \\ 
 BOSS z2 NGC & $\{0.4 - 0.6\}$ & $0.51$ & $500 \ 872$ & $3.1$ & $\{0.2, \ 0.2, \ 0.15 \}$ \\
 BOSS z2 SGC & $\{0.4 - 0.6\}$ & $0.51$ & $185 \ 498$ & $1.1$ & $\{0.2, \ 0.2, \ 0.15 \}$ \\ 
 LRGpCMASS NGC & $\{0.6 - 1.0\}$ & $0.698$ & $255 \ 741$ & $2.72 \ (\rm{NGC + SGC})$ & $\{0.2, \ 0.2, \ 0.15 \}$ \\ 
 LRGpCMASS SGC & $\{0.6 - 1.0\}$ & $0.698$ & $121 \ 717$ & $2.72 \ (\rm{NGC + SGC})$ & $\{0.2, \ 0.2, \ 0.15 \}$ \\ 
 eBOSS QSO NGC & $\{0.8 - 2.2\}$ & $1.52$ & $218 \ 209$ & $0.35$ & $\{0.2, \ 0.2, \ 0.15 \}$ \\ 
 eBOSS QSO SGC & $\{0.8 - 2.2\}$ & $1.52$ & $125 \ 499$ & $0.18$ & $\{0.2, \ 0.2, \ 0.15 \}$ \\ 
 \midrule
 Nseries & $\{0.43 - 0.7\}$ & $0.56$ & $\sim 660 \ 000$ & $3.67 \ (\times 84)$ & $\{0.2, \ 0.2, \ 0.15 \}$ \\
\end{tabular}
\caption{Full list of clustering measurements which constitute our baseline full-shape data vector, and mock catalogues used for model verification. Values like $z_{\rm{eff}}$ and $V_{\rm{eff}}$ have been taken directly from the Fourier space analysis papers, and may be computed using different techniques between surveys. Note that the values quoted for our Nseries sample correspond to the properties of a single mock realisation.}
\label{tab:DatasetList}
\end{table*}

\begin{figure*}
    \centering
        \includegraphics[width=18cm]{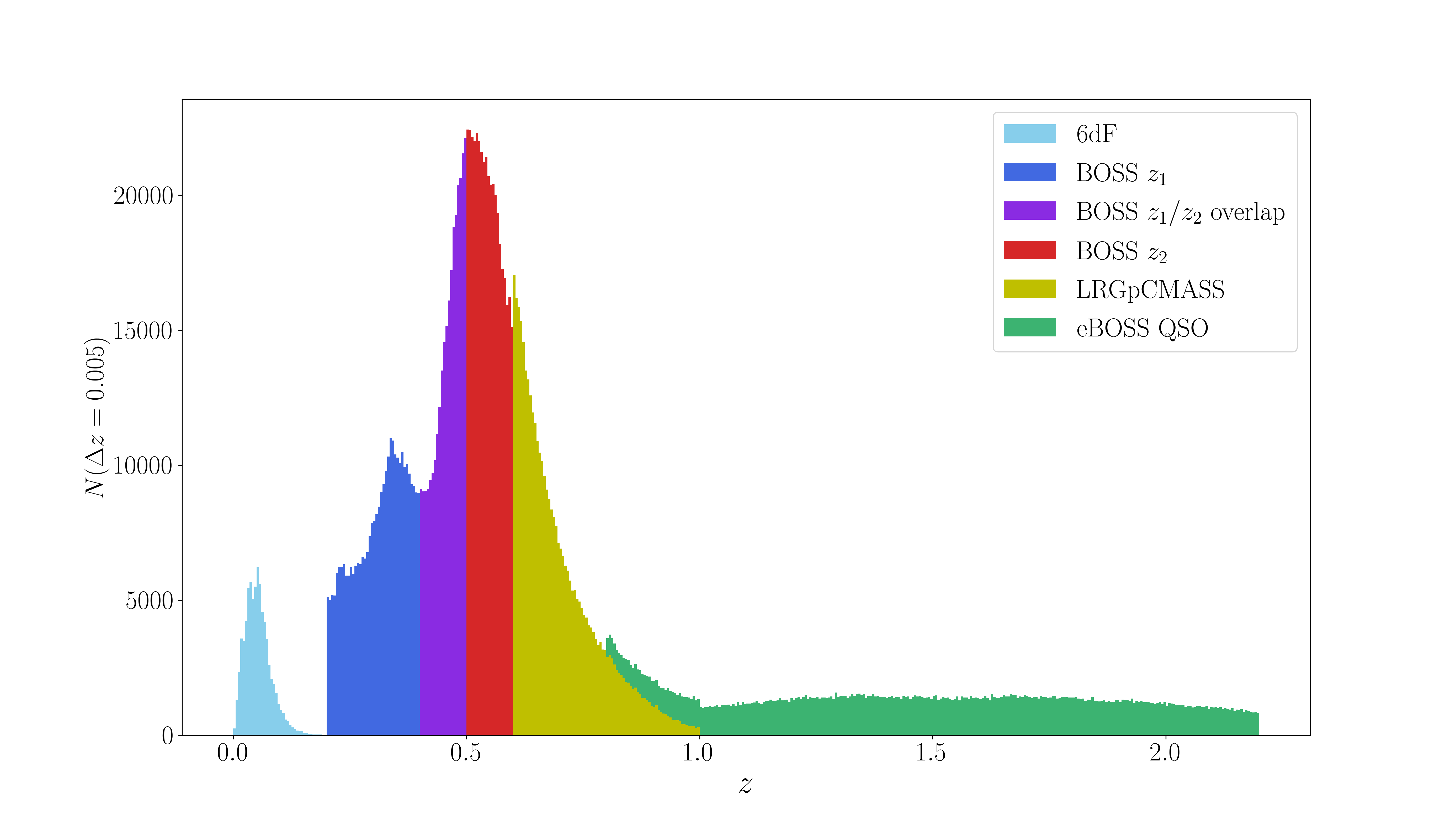}
        \caption[]{Redshift distribution of catalogues used in this analysis, counted over bins of width $\Delta z = 0.005$. While the bulk of our redshifts are within the $z = \{0.2 - 0.8\}$ regime, the inclusion of nearby (6dFGS) and very distant (eBOSS QSO) sources help address degeneracies in the full-shape modelling. The purple region corresponds to the overlapping samples of the BOSS $z_1$ and $z_2$ catalogues--- While this overlap does not introduce any additional sources over total $z = \{0.2 - 0.6 \}$ region, it does increase the number of galaxy pairs available for modelling.}
        \label{fig:RedshiftHistogram}
\end{figure*}

\subsection{Power Spectrum Measurements}
\label{sec:PowerSpec}

Throughout this work, we use the \textsc{PyBird} code to simultaneously constrain clustering measurements from the 6dFGS, BOSS, and eBOSS catalogues
\citep{6dF_paper, Alam_2017, eboss_2020}. The BOSS and eBOSS catalogues are composed of measurements made in two non-overlapping sky patches, referred to as the Northern and Southern Galactic Caps (hereafter NGC and SGC). Following the convention of \cite{Alam_2017}, we further divide the BOSS sample into two overlapping redshift bins, $0.2<z_1<0.5$, and $0.4<z_2<0.6$, for a total of four BOSS clustering measurements. Note that the clustering statistics for these four BOSS samples were computed directly from observational catalogues by us, using the \textsc{Nbodykit} toolkit \citep{nbodykit2018}. In line with the approach of \cite{GilMarin2020}, we combine the high-$z$ BOSS CMASS and eBOSS LRG samples, yielding two ``LRGpCMASS'' samples at $z_{\textrm{eff}} = 0.698$. Finally, we include the NGC and SCG clustering measurements from the eBOSS QSO catalogue, at $z_{\textrm{eff}} = 1.52$. This combination of cuts yields a total of 9 unique clustering measurements (summarised in Table \ref{tab:DatasetList}, and visualised in Figure \ref{fig:RedshiftHistogram}), which form the joint data vector for our full-shape analysis In order to verify the accuracy and reliability of our analysis pipeline (across both flat and curved cosmological models), we also constrain cosmology from the mean of 84 Nseries N-body mock catalogues (hereafter referred to as the ``Nseries'' sample), once again computed using \textsc{Nbodykit}. 

When computing clustering statistics from these observational catalogues, a fiducial cosmology must be assumed in the conversion of redshifts to comoving distances. To test how this choice could impact full-shape curvature constraints, we compute clustering statistics from the Nseries mock catalogues using the three fiducial cosmologies given in Table \ref{tab:fid_cosmo_list}. The power spectra used in our mock validation runs (the fits of Section \ref{sec:MockVerification}) are conducted using our flat fiducial cosmology. We then repeat our analysis of the \textsc{Nseries} mock catalogues in Section \ref{sec:Fiducial_Cosmology_Section}, using power spectra which employ closed $\Omega_k = +0.1$) and open ($\Omega_k = -0.1$) fiducial cosmologies. These spatially curved fiducial cosmologies are otherwise identical to our flat fiducial cosmology, with $\Omega_k$ set by fixing $\Omega_{\Lambda}$. 

\begin{table}
\centering
\begin{tabular}{lccccc}
\hline
 & $\Omega_{m}$ & h & $\Omega_{b}$ & $\Omega_{\Lambda}$ & $\Omega_{k}$ \\ \hline
\textbf{Flat} & 0.31 & 0.676 & 0.0481 & \textbf{0.693} & \textbf{0} \\
\textbf{Closed} & 0.31 & 0.676 & 0.0481 & \textbf{0.593} & \textbf{0.1} \\
\textbf{Open} & 0.31 & 0.676 & 0.0481 & \textbf{0.793} & \textbf{-0.1} \\ \hline
\end{tabular}
\caption{Fiducial cosmologies used to analyse Nseries mocks. Our $\Omega_{k} = 0$ cosmology corresponds to the fiducial cosmology employed in \protect\cite{Alam_2017}, where we modify $\Omega_k$ by fixing $\Omega_{\Lambda}$ from this baseline. In principle, the offset between our flat fiducial cosmology from the underlying cosmology of the Nseries sample ($\Omega_m = 0.286$) exposes our analysis to the systematics described in Section \ref{sec:fiducial_cosmology}. In practice, the effect of this flat fiducial offset is known to be negligible for standard analyses, whereas the same is not a priori known for different choices of $\Omega_k$.}
\label{tab:fid_cosmo_list}
\end{table}

A number of investigations have already used full-shape measurements to place constraints on curvature from different combinations of the BOSS data. Notably, \cite{Chudaykin_2021} recover $\Omega_k = -0.043 \pm 0.036$ from joint fits which include power spectrum measurements derived from the DR12 $z_1$ and $z_3$ bins, along with post reconstruction BAO and supernovae information. Similarly, \cite{Vagnozzi2020} combine power spectrum measurements from the DR12 CMASS sample ($0.43 < z < 0.7$) with Planck information, recovering constraints consistent with flatness to $\sim 0.2 \%$. Our combination of 6dFGS, BOSS, and eBOSS data represents an almost uninterrupted series of clustering measurements across a large fraction of cosmic history, providing an unprecedented level of constraining power from full-shape analyses. We find this remarkable completeness across expansion history helps insulate EFT constraints against parameter degeneracies which emerge in full-shape only constraints of extended cosmological models (particularly the curved $\Lambda$CDM fits conducted in this work). 

Importantly, this completeness does require the use of the overlapping $z_1$ and $z_2$ redshift bins from BOSS, introducing non-trivial correlations across these samples. To account for this effect, we compute the full cross-covariance matrix for our overlapping BOSS samples using the publically available MultiDark Patchy mock catalogues, provided as part of BOSS DR12 \citep{Kitaura_2016, RodriguezTorres_2016}. All other sky-chunks in our analysis are treated as independent, and combined through the simple addition of their respective likelihoods. It should be noted that while there is a very small overlap between our LRGpCMASS and QSO samples across $z = \{0.8 - 1 \}$ (as seen in Figure \ref{fig:RedshiftHistogram}), this region does not significantly contribute to the two sets of measurements. \footnote{\cite{Stoecker_2021} found a cross-correlation coefficient between the DR14 eBOSS LRG and QSO BAO measurements of 0.29, but this was found to be $<0.1$ for the DR16 LRGpCMASS and QSO samples as the overlapping cosmological volume constitutes a far smaller fraction of the total volume for these latter two samples \citep{eboss_2020}.} Finally, we de-bias our estimate of the numerical inverse of the covariance matrix for all sky-chunks using the correction proscribed in \cite{Hartlap2007}.

\subsection{Analysis Methodology}

\subsubsection{Summary Statistics}

In cosmology, we are generally interested in measuring the posterior distribution of some ensemble of variables, to update their constraints in light of some new information or measurement. This is succinctly encoded in Bayes theorem,
\begin{equation}
    P(\theta| D) = \dfrac{P(D|\theta) P(\theta)}{P(D)} = \dfrac{\mathcal{L}(\theta) \pi(\theta)}{\mathcal{Z}},
    \label{eq:BayesTheorem}
\end{equation}
for a posterior P($\theta$|D), likelihood $\mathcal{L}$, prior $\pi$, and evidence $\mathcal{Z}$. 

While most of our attention is generally focused on P($\theta$|D) for the purposes of model fitting, the evidence plays a central role in model \textit{comparison} by quantifying the performance of a model in the light of the data \citep{Trotta_2008}. In particular, we can compare the compatibility of two models with the same a priori likelihood through the ratio of their evidences. This technique has already been employed to shed light on the possible curvature tension, with \citet{Handley2019} assigning 50:1 Bayesian betting odds against flatness from Planck 2018 CMB information. 
The Bayesian evidence can also be used to determine the congruity of multiple datasets under combination, through the calculation of their Bayes factor,

\begin{equation}
    R = \dfrac{\mathcal{Z}_{AB}}{\mathcal{Z}_{A} \mathcal{Z}_{B}},
\end{equation}

where $\mathcal{Z}_{AB}$ denotes the evidence derived from the product of likelihoods $\mathcal{L}_{A}$ and $\mathcal{L}_{B}$. While the Bayes factor is intrinsically prior dependent, \cite{Lemos_2019} demonstrate how this sensitivity can be minimised using the change in Kullback-Leibler divergence ($\mathcal{D}$), yielding the ``suspiciousness statistic'',
\begin{equation}
    S = \frac{R}{I}, \ \ \textrm{log}(I) = \mathcal{D}_{1} + \mathcal{D}_{2} - \mathcal{D}_{12}.
\end{equation}


In order to more broadly understand how full-shape measurements inform curvature constraints, we measure $\mathcal{Z}$ for a range of dataset/model combinations, using the evidence ratio to determine whether full-shape information alone exhibits a preference for the inclusion of curvature in model evaluation. We then make use of the Bayes factor to determine the compatibility of full-shape information with Planck CMB data. When we quantify this internal consistency, we follow the approach of \cite{Handley2019} and convert the suspiciousness statistic into a ``tension probability'' ($p$), using the survival function of the chi-squared distribution,
\begin{equation}
    p = \int^{\infty}_{d - 2 \textrm{log(S)}} \chi^2_d(x) dx,
\end{equation}
\begin{equation}
    d = d_A + d_B - d_{AB}.
\end{equation}
$d_A$, $d_B$, and $d_{AB}$ correspond to the Bayesian model dimensionalities for our individual and joint datasets respectively, defined as
\begin{equation}
    d = 2 * (\langle \textrm{log}(\mathcal{L}) ^2 \rangle - \langle \textrm{log}(\mathcal{L}) \rangle ^2 )
\end{equation}
We also quote this tension probability as a sigma value, calibrated as
\begin{equation}
    \sigma = \sqrt{2} \textrm{Erfc}^{-1} (p).
\end{equation}

\subsubsection{Sampling Strategy}

\label{sec:SamplingStrategy}


In order to efficiently sample our posterior distribution, we optimize parameter jumps using adaptive Metropolis-Hastings sampling through \textsc{MontePython} \citep{brinckmann_2018}. When computing evidences for model comparison, we use \textsc{MontePython} as an interface with the nested sampling algorithm \textsc{Polychord} \citep{Handley_2015}. For each step along our cosmological parameter space, we use the Boltzmann code \textsc{CLASS} to compute linear power spectra for each unique clustering signal, evaluated at their respective $z_{\textrm{eff}}$. We then use \textsc{PyBird} to compute the one-loop non-linear power spectrum from these linear signals, subject to our EFT bias parameters. To replicate the survey geometry and selection effects present in observational data, each model power spectrum must be convolved with their corresponding window function. When fitting the full ensemble of observational catalogues given in Table \ref{tab:DatasetList}, this yields 9 model $P(k)$ which may be directly compared to $P(k)$ from our observed catalogues.

For our full-shape only posterior evaluations, we directly sample up to 6 cosmological parameters \{ln($10^{10} A_s$), $h$, $\Omega_{\textit{cdm}}$, $\Omega_b$, $n_s$, $\Omega_k$\}, with 2 additional free EFT bias parameters for each clustering sample \{$b_1$, $c_2$\}. We begin by employing broad, flat priors over almost all of our cosmological and bias parameters,
\begin{equation}
\begin{split}
    \rm{ln}(10^{10} A_s) = \{1, 4\} \hspace{2cm} h = \{0.5, 0.85\} \\
    \Omega_{\textit{cdm}} h^2 = \{0.05, 0.25\} \hspace{1.4cm} \Omega_{k} = \{-0.25, 0.25\} \\
    b_1 = \{0, 4\} \hspace{2.3cm} c_2 = \{-4, 4\}
\end{split}
\end{equation}

\begin{table*}[t]
\small
\centering
\renewcommand{\arraystretch}{2.0} 
\setlength{\tabcolsep}{3.0pt}
\begin{tabular}{@{}lllllll@{}}
\toprule
 & $\rm{ln}(10^{10} A_{\textit{s}})$ & $h$ & $\Omega_{\textit{cdm}} h^2$ & $\Omega_{\textit{m}}$ & $\Omega_k$ \\ \midrule
 \textbf{Truth:} & \textbf{3.066} & \textbf{0.7} & \textbf{0.117} & \textbf{0.286} & \textbf{0} \\ \midrule 
 Nseries, flat $\Lambda$CDM, std cov & $3.04^{+0.21}_{-0.22} \ (3.12)$ & $0.699^{+0.017}_{-0.017} \ (0.694)$ & $0.114^{+0.009}_{-0.009} \ (0.110)$ & $0.281^{+0.016}_{-0.015} \ (0.275)$ & -  \\
 Nseries, flat $\Lambda$CDM, red cov ($25$) & $3.05^{+0.05}_{-0.05} \ (3.05)$ & $0.700^{+0.006}_{-0.006} \ (0.701)$ & $0.116^{+0.003}_{-0.003} \ (0.116)$ & $0.284^{+0.004}_{-0.004} \ (0.284)$ & - \\ \midrule
 Nseries, curved $\Lambda$CDM, std cov & $2.96^{+0.39}_{-0.42} \ (3.17)$ & $0.702^{+0.031}_{-0.030} \ (0.693)$ & $0.114^{+0.008}_{-0.008} \ (0.111)$ & $0.275^{+0.026}_{-0.025} \ (0.279)$ & $-0.03^{+0.11}_{-0.11} \ (0.02)$ \\
 Nseries, curved $\Lambda$CDM, red cov (25) & $3.08^{+0.10}_{-0.11} \ (3.04)$ & $0.699^{+0.009}_{-0.010} \ (0.701)$ & $0.116^{+0.003}_{-0.003} \ (0.116)$ & $0.285^{+0.007}_{-0.006} \ (0.282)$ & $0.01^{+0.03}_{-0.03} \ (-0.01)$ \\ \midrule
\end{tabular}
\caption{Constraints recovered through fitting our high quality mock mean data against $\Lambda$CDM models with fixed curvature ($\Omega_k$ = 0), and models which permit free variations in curvature. with the maximum a posteriori cosmology for each model given in brackets. Our model is able to robustly recover the truth cosmology of the Nseries reference sample, across both standard and reduced covariance fits. While the inclusion of $\Omega_k$ as a free parameter does affect our constraining power on $\rm{ln}(10^{10} A_{s})$ and $h$, this does not work to systematically shift our model away from the underlying truth cosmology. This consistent preference for our truth cosmology serves to emphasise the robustness of our EFT pipeline when used to constrain $\Omega_k$.}
\label{tab:MockMeanFits}
\end{table*}

While clustering measurements can weakly constrain $\Omega_{b}$ (through the relative ratio of the power spectrum peaks), information from big bang nucleosynthesis (BBN) measurements offer robust, CMB-independent constraints on the baryonic matter density. We use the measurements provided in \cite{Cooke_2018}, to place a Gaussian prior on our fits of $100 \times \Omega_{b} h^2 = 2.235 \pm {0.049}$. While the free-streaming of massive neutrinos does impact the shape of the matter power spectrum, for realistic neutrino masses this effect is firmly subdominant to our parameters of interest. For simplicity, we therefore assume a cosmological model with three massless neutrino species ($N_{\textrm{eff}} = 3.046$) across all full-shape fits. 

In principle, the full-shape of the power spectrum can be used to directly constrain the tilt of the primordial power spectrum ($n_s$), however such constraints are an order of magnitude weaker than CMB measurements \citep{Chudaykin_2020}. Across our full-shape only constraints, we therefore undertake each analysis twice--- once using a spectral tilt fixed to $n_s = 0.965$ \citep[taken from][]{Planck_2018}, and once with $n_s$ as a free parameter. The only exception to this is our model verification, where $n_s$ is fixed to the truth values of the Nseries catalogue ($n_s = 0.96$). Finally, when running fits to clustering measurements (across both mock and observed data-sets), we choose to constrain the monopole, quadrupole, and hexadecapole ($P_0$, $P_2$, $P_4$) signals from $k_{\rm{min}} = \{0.01, 0.01, 0.01\}$ to $k_{\rm{max}} = \{0.2, 0.2, 0.15\}$, with the exception of the 6dFGS measurements (which are fit from $k_{\rm{min}} = \{0.01, 0.01, 0.01\}$ to $k_{\rm{max}} = \{0.15, 0.15, 0.1\})$.

\section{Full-Shape Model Tests}
\label{Sec:Results}

\subsection{Mock Validation}
\label{sec:MockVerification}

\begin{figure}
        
        
        \includegraphics[width=8.5cm]{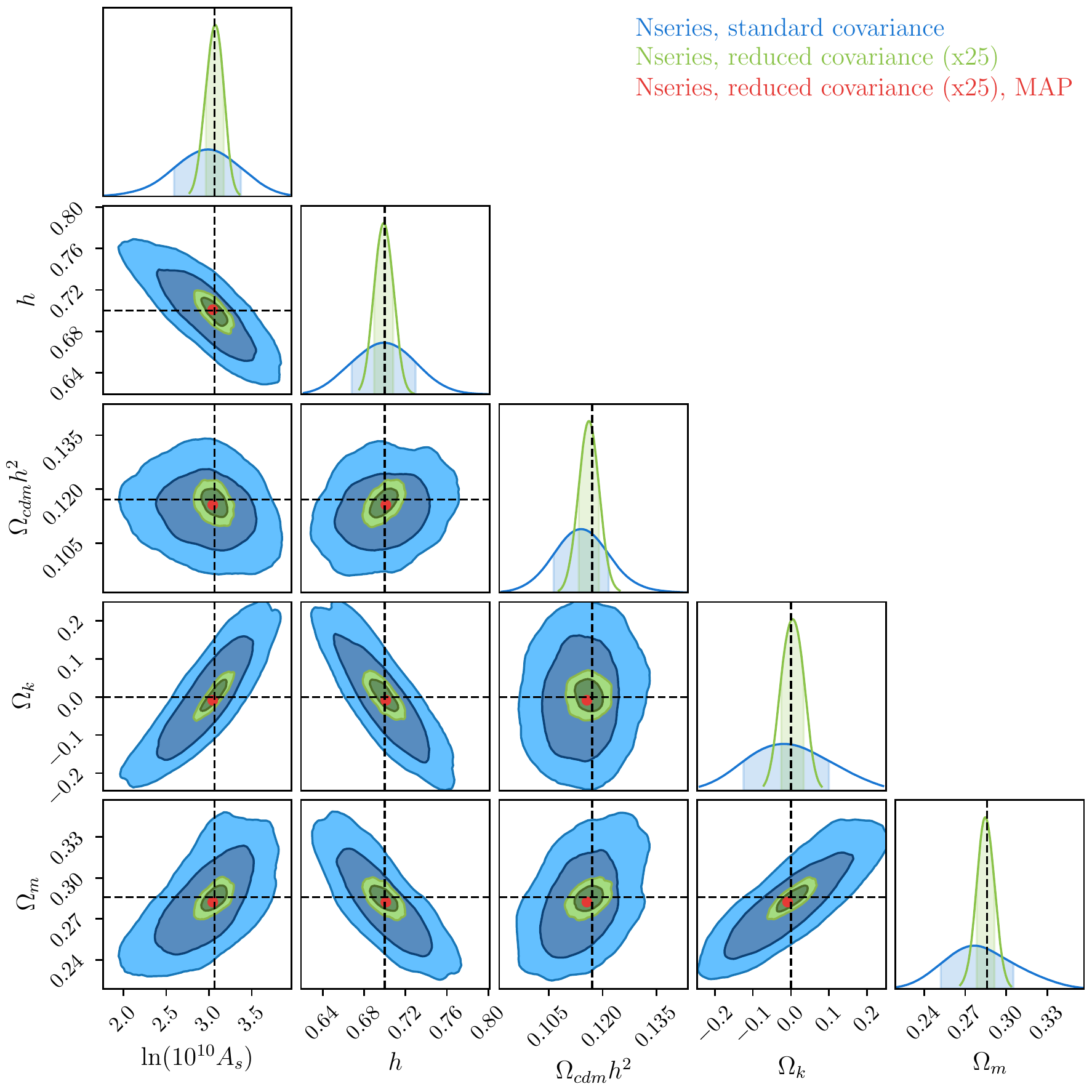}
        \caption[]{Posteriors recovered from analysis of our Nseries mock reference sample in a curved $\Lambda$CDM framework, using our standard and reduced amplitude covariance matrix (with the maximum a posterior cosmology for our reduced covariance fit given in red). Notably, the inclusion of curvature as a free parameter does not shift our posterior mean or a posteriori point away from the truth value of the Nseries mock catalogues. As the errors assigned to our reference sample are reduced our model remains unbiased, with both the posterior mean and maximum a posteriori cosmology lining up with the truth value of our reference catalogue.}
         \label{fig:mock_mean_flat_and_curved_posterior}
\end{figure}

In order to test the reliability of our fitting pipeline, we begin by constraining the mean of 84 pre-reconstruction Nseries mock catalogues. The Nseries mocks are full N-body simulations which replicate the observational pattern of the NGC $0.43 < z < 0.7$ sample from the BOSS DR12 catalogue, generated using a fixed $\Lambda$CDM cosmology (defined with $\textrm{ln}10^{10} A_s$ = 3.066, $h$ = 0.7, $\Omega_m$ = 0.286, $\Omega_b$ = 0.047, $\Omega_{k}$ = 0.0). In this verification stage we constrain the mean of 84 individual Nseries mock realisations (propagated using the flat fiducial cosmology of Table \ref{tab:fid_cosmo_list}), creating an effectively noiseless ``reference sample'' to establish the accuracy of results recovered from our analysis scheme. In order to rigorously examine the impact of potential systematic errors, we repeat our analysis using a standard covariance matrix (designed to replicate the expected errors from the real measurement in the same volume), and a covariance matrix which has been reduced in amplitude by a factor of 25. To compute the standard errors for this reference sample we follow the approach of \cite{GilMarin2020}, and use the variance in 2048 MultiDark Patchy Mock catalogues over the above redshift range (with an additional $10 \%$ rescaling factor to account for veto effects applied to the MultiDark mocks, but not the Nseries catalogues). We provide the confidence interval and best-fit from constraints of this high fidelity sample in Table \ref{tab:MockMeanFits}, with the full posterior from fits to curved models provided in Figure \ref{fig:mock_mean_flat_and_curved_posterior}.

We find our analysis pipeline is able to recover the underlying cosmology of the Nseries sample to a high degree of fidelity, across models with both fixed and varying curvature. As the errors on this reference sample are reduced, our model distinctly converges around the underlying Nseries cosmology (most reassuringly, recovering $\Omega_{k, \textrm{true}} = 0$). A notable degeneracy between $\Omega_k$ and the scalar amplitude parameter $A_s$ can be identified in our full posterior distribution, significantly weakening our constraining power for curved models. Despite the presence of this degeneracy, the fidelity of our posterior and best-fit across all validation fits strongly indicates the full-shape model employed in this work provides unbiased measurements of flat cosmologies, and can be applied to observation.

\begin{figure}
        \includegraphics[width=8.66cm]{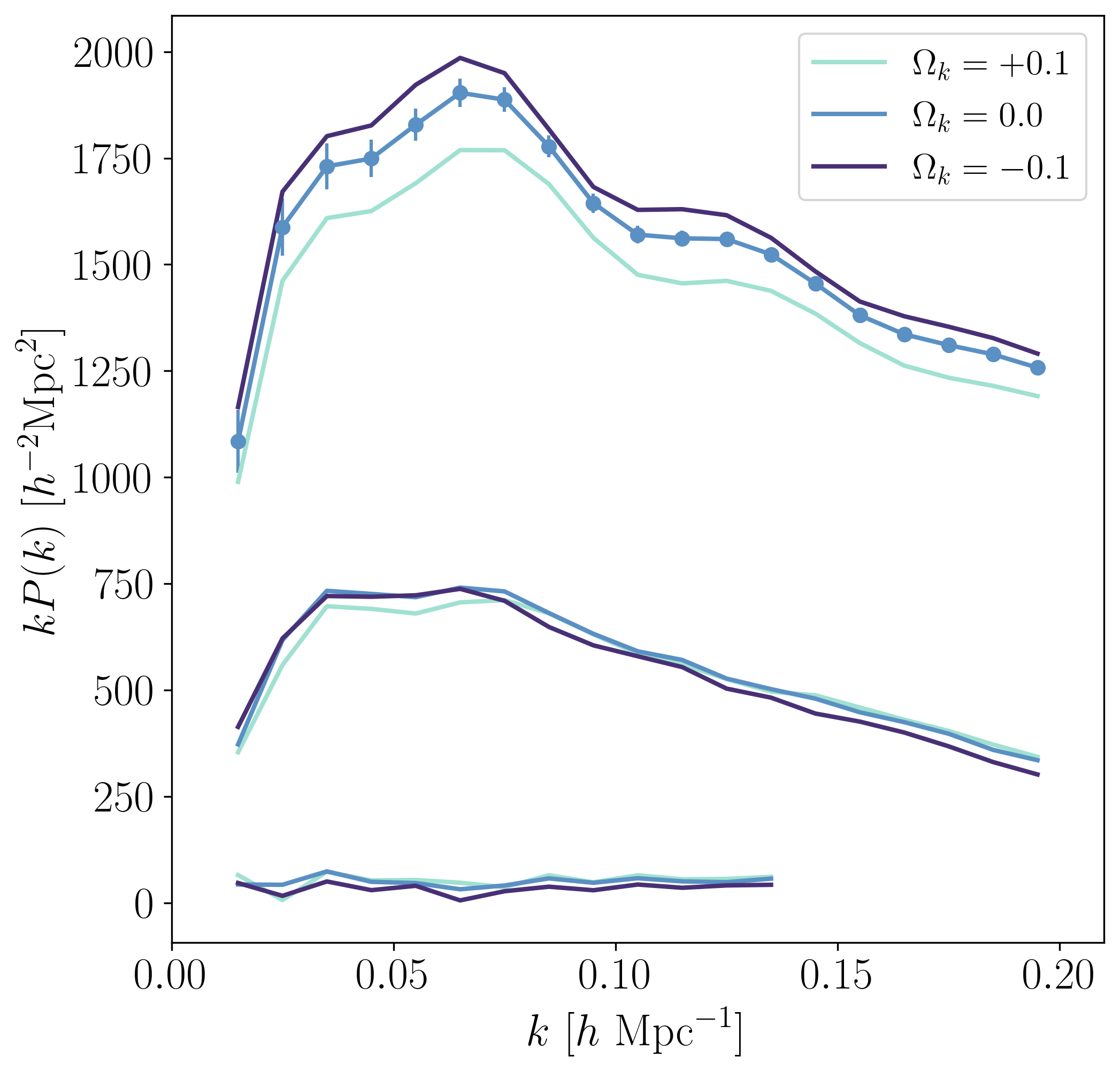}
        \caption[]{The Monopole (upper), Quadrupole (middle) and Hexadecapole (lower) signals generated from propagating the Nseries mocks using the three fiducial cosmologies given in Table \ref{tab:fid_cosmo_list}. Note that we also include the standard $P_0$ errors on our $\Omega_{\textrm{k, fid}} = 0$ sample for reference. Non-flat ($\Omega_{\textrm{k, fid}} = \pm 0.1$) cosmologies are defined by fixing $\Omega_m$, $\Omega_b$, and $h$ to the default values of our flat fiducial cosmology, where $\Omega_{\Lambda}$ is varied to achieve the desired $\Omega_k$. The impact of our non-flat fiducial cosmology is to uniformly dilate the matter power spectrum across $k$, resulting in a visible dilation of the BAO feature. The modification of our redshift-distance relationship also visibly modulates the overall amplitude of the Monopole signal, well in excess of the measurement errors.}
        \label{fig:Nseries_varying_fiducial}
\end{figure}

\begin{figure*}
\centering
        
        
        \includegraphics[width=12cm]{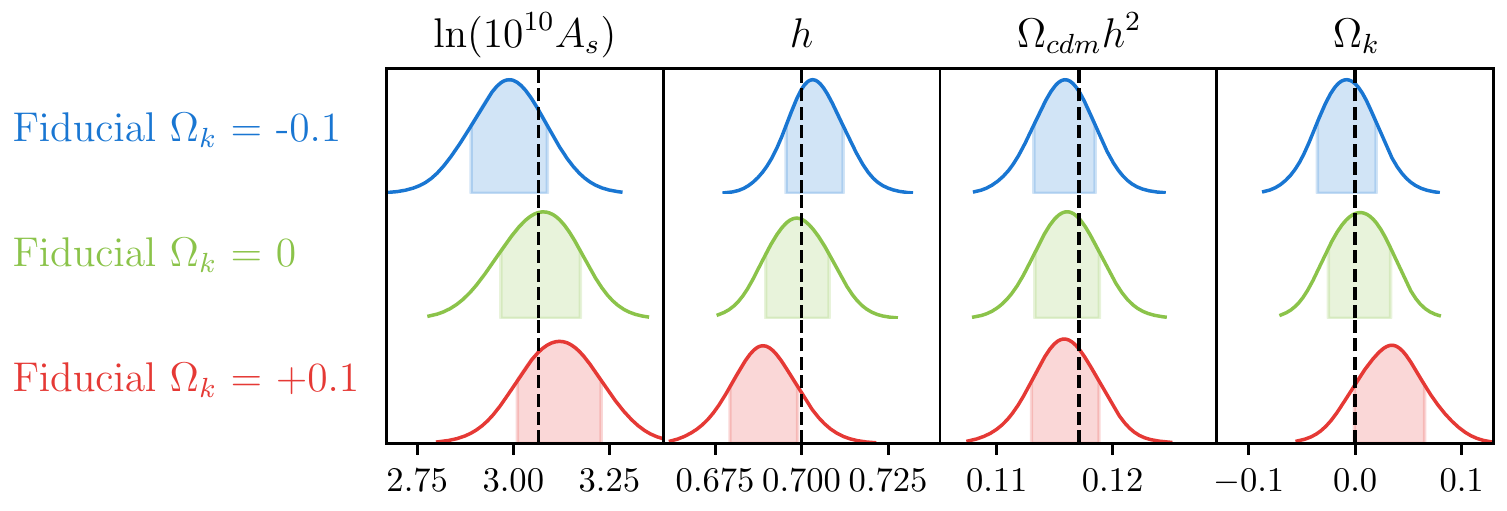}
        \caption[]{Reduced amplitude covariance fits to the mean of 84 Nseries mock catalogues using the three fiducial cosmologies of Table \ref{tab:fid_cosmo_list}, with a dotted line corresponding to the posterior mean of our flat fiducial sample for reference. We find the use of a fiducial cosmology which substantially deviates from the underlying cosmology of our sample results in a small bias on $\Omega_k$ fits from our flat baseline, but which is subdominant to the statistical variance of a standard measurement ($\sim 0.2 \sigma$).}
        \label{fig:fid_cosmo_test_contour}
\end{figure*}

\subsection{Impact of Fiducial Cosmology}

\label{sec:Fiducial_Cosmology_Section}

With the accuracy of our full-shape pipeline validated against an effectively noiseless mock-mean reference catalogue, we now consider the role that our choice of fiducial cosmology plays in full-shape constraints of $\Omega_k$, in line with Section \ref{sec:fiducial_cosmology}. We fully recompute the clustering statistics for our suite of 84 Nseries mock catalogues using two extremely curved fiducial cosmologies (with  $\Omega_{k\rm{, fid}} = \pm 0.1$, given in Table \ref{tab:fid_cosmo_list}). It should be noted that the offset between the underlying cosmology of our sample, and the chosen fiducial cosmology of propagation, is much larger than the expected offset in realistic fits. Indeed, this test was designed to be as sensitive as possible to any residual effects which may arise from the use of a non-flat fiducial cosmology in analysis, which can be traced back to model assumptions or approximations. In line with this philosophy, we also use the reduced amplitude covariance matrix of Section \ref{sec:MockVerification} throughout fitting, reducing the final width of our measurement errors by a factor of 5. When recomputing clustering measurements from this sample, we also recompute the number density, FKP weights, and Poisson shot noises assigned to these samples, yielding the clustering signals in Figure \ref{fig:Nseries_varying_fiducial}. Additionally, we fully recompute the window function applied to our model power spectra, using random catalogues which have been propagated using the same (curved) fiducial cosmology as our mock data. Using these clustering signals to constrain our curved $\Lambda$CDM model model (broadly repeating the technique of Section \ref{sec:MockVerification}), we recover the posterior distributions provided in Figure \ref{fig:fid_cosmo_test_contour}.


After fully propagating our modified fiducial cosmology across all stages of analysis, we recover a modest scatter in parameters which  modulate the amplitude of our clustering signal. Notably, the use of a fiducial cosmology which is offset from our underlying sample by $\Omega_{\textrm{k, fid}} = +0.1$ yields a small offset in $\Omega_k$--- Importantly, this effect is sub-dominant to the statistical error assigned by our unmodulated covariance matrix ($\sim 0.2 \sigma$). Fiducial cosmological assumptions are therefore unlikely to represent a major source of error in present day full-shape constraints of curvature. As future galaxy redshift surveys provide unprecedented improvements in scale and accuracy, the constraints offered by full-shape measurements are likely to grow in importance as stand alone measurements. While outside the scope of this work, it is important to consider whether assumptions in fiducial cosmology (and indeed assumptions of flatness in general) will continue to remain negligible with increasing precision. 

\section{EFTofLSS and the "Curvature Tension"}

\begin{figure*}
    \centering
        
        \includegraphics[width=8.8cm]{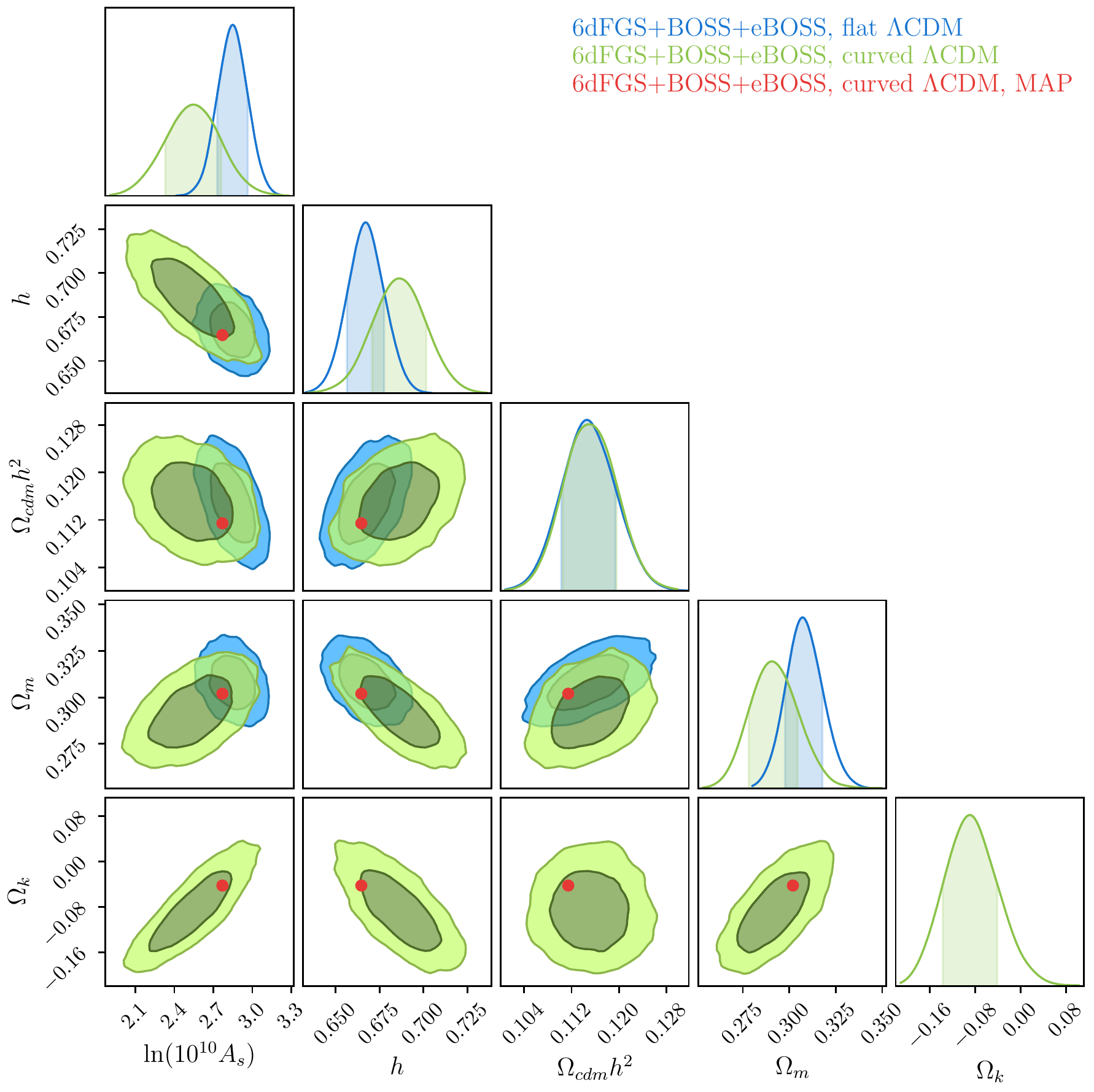}
        \includegraphics[width=8.8cm]{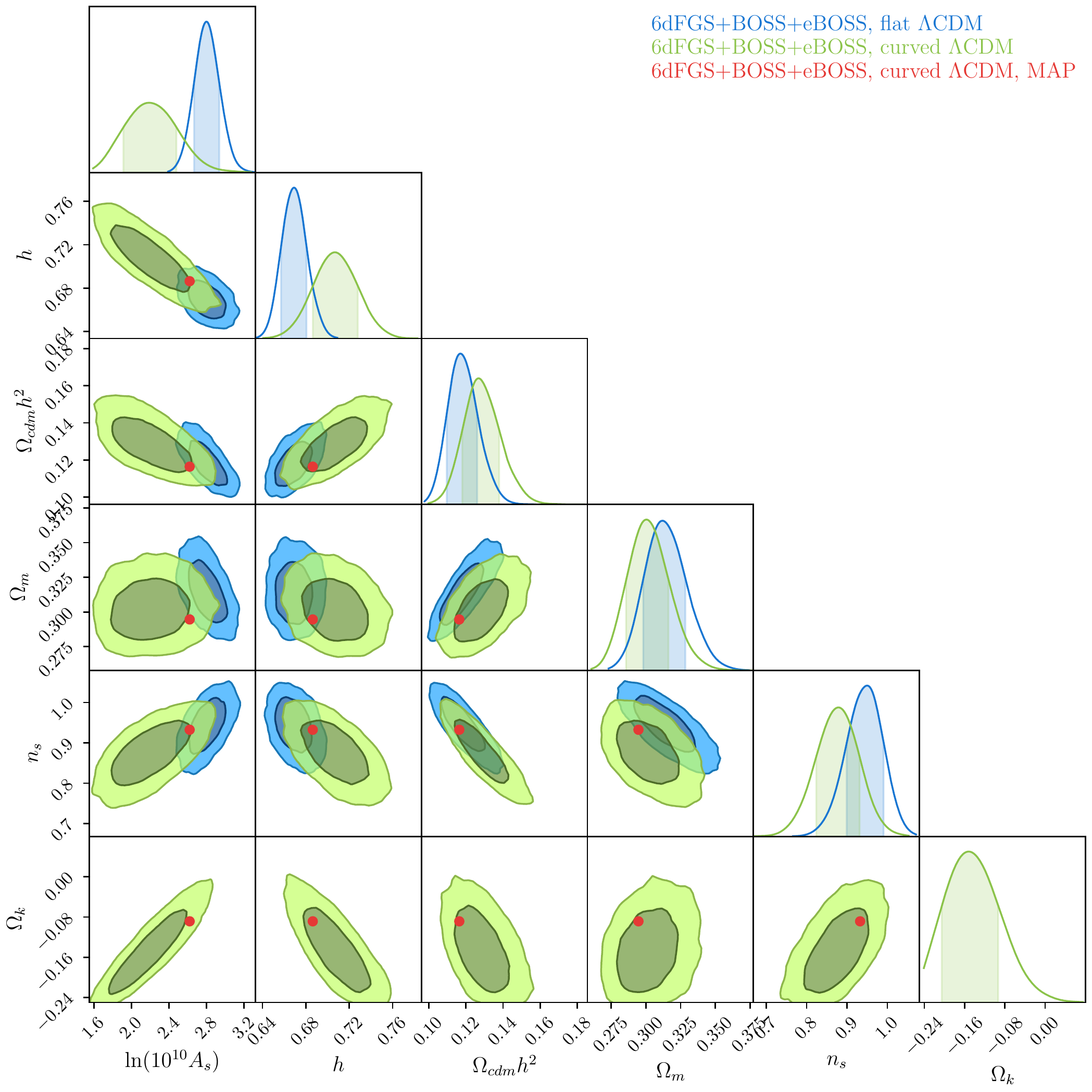}
        \caption[]{Posteriors recovered from joint 6dFGS+BOSS+eBOSS constraints in both flat and curved $\Lambda$CDM models, when $n_s$ is fixed to the Planck 2018 value (left) and when $n_s$ is allowed to vary (right). Consistent with the findings of our mock verification, the inclusion of $\Omega_k$ substantially weakens our constraints on  $\rm{ln}(10^{10} A_s)$ and $h$, stretching these posteriors along their degeneracy direction. When $n_s$ is allowed to vary in addition to $\Omega_k$, the reduced constraining power on $\Omega_{\textit{cdm}} h^2$ further compounds this degeneracy, offsetting marginalised parameter constraints from flat models. Interestingly, across all tests our maximum a posterori cosmologies are consistently skewed along this degeneracy direction, yielding best-fits which are substantially closer to flatness than the posterior mean.}
        \label{fig:TrueDataPosterior}
\end{figure*}

\begin{table*}
\footnotesize
\centering
\renewcommand{\arraystretch}{2.0} 
\setlength{\tabcolsep}{1.5pt}
\begin{tabular}{@{}lllllllll@{}}
\toprule
 & $\rm{ln}(10^{10} A_{\textit{s}})$ & $h$ & $\Omega_{\textit{cdm}} h^2$ & $\Omega_{\textit{m}}$ & $\Omega_k$ & $n_s$ & $2*\rm{log}(\mathcal{L})$ \\ \midrule
 Flat, fixed $n_s$ & $2.85^{+0.11}_{-0.12} \ (3.03)$ & $0.667^{+0.011}_{-0.011} \ (0.672)$ & $0.114^{+0.005}_{-0.004} \ (0.115)$ & $0.307^{+0.010}_{-0.011} \ (0.304)$ & -  & - & $367.2$ \\
 Curved, fixed $n_s$ & $2.55^{+0.21}_{-0.22} \ (2.77)$ & $0.686^{+0.015}_{-0.016} \ (0.665)$ & $0.115^{+0.004}_{-0.005} \ (0.111)$ & $0.291^{+0.014}_{-0.013} \ (0.302)$ & $-0.089^{+0.049}_{-0.046} \ (-0.042)$ & - &  $366.3$ \\
 \midrule
 Flat, varying $n_s$ & $2.80^{+0.14}_{-0.13} \ (2.97)$ & $0.669^{+0.012}_{-0.011} \ (0.668)$ & $0.117^{+0.009}_{-0.008} \ (0.114)$ & $0.312^{+0.017}_{-0.014} \ (0.304)$ & -  & $0.950^{+0.04}_{-0.051} \ (0.972)$ & $367.1$ \\
 Curved, varying $n_s$ &  $2.19^{+0.29}_{-0.28} \ (2.62)$ & $0.707^{+0.021}_{-0.021} \ (0.686)$ & $0.127^{+0.011}_{-0.009} \ (0.116)$ & $0.300^{+0.016}_{-0.014} \ (0.295)$ & $-0.152^{+0.059}_{-0.053} (-0.089)$  & $0.878^{+0.053}_{-0.055} \ (0.932)$ & $364.8$
 \\ \midrule
\end{tabular}
\caption{$68\%$ confidence interval and best fit provided from analysis of our combined 6dFGS+BOSS+eBOSS sample, across both fixed and varying $n_s$/$\Omega_k$. When analysed in isolation, our suite of pre-reconstruction full-shape measurements exhibit a $\sim 2 \sigma$ preference for curvature across all fits. Interestingly, while our curved models generally perform better than their flat counterparts (as measured by $\rm{log}(\mathcal{L})$), the maximum a posteriori point in both fits is skewed from the posterior mean by $\sim 1 \sigma$.}
\label{tab:6dF_BOSS_eBOSS_table}
\end{table*}


\subsection{6dFGS+BOSS+eBOSS Fits}

\label{sec:TrueDataResults}

\begin{figure*}
    \centering
        
        
        \includegraphics[width=5.8cm]{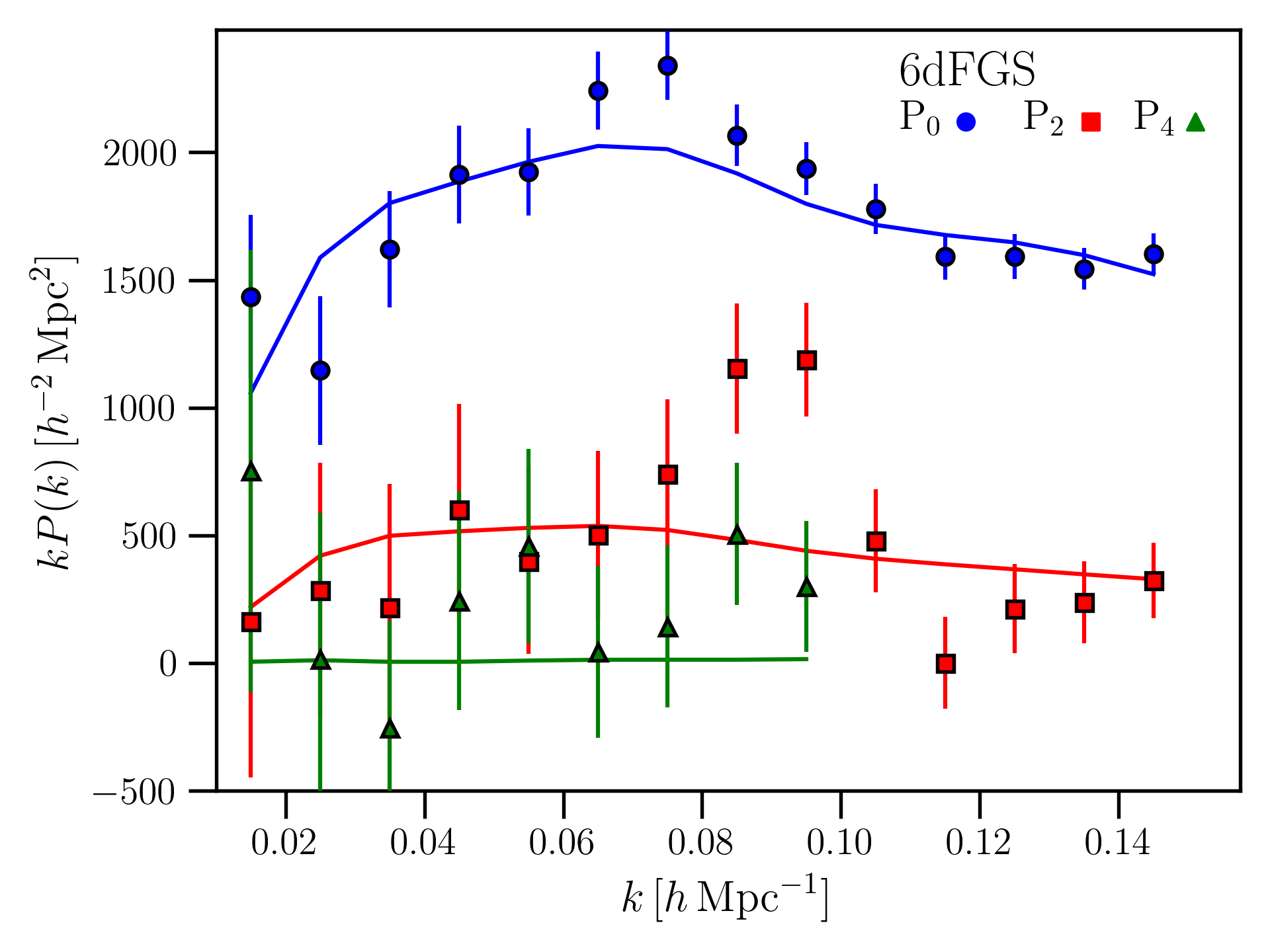}
        \includegraphics[width=5.8cm]{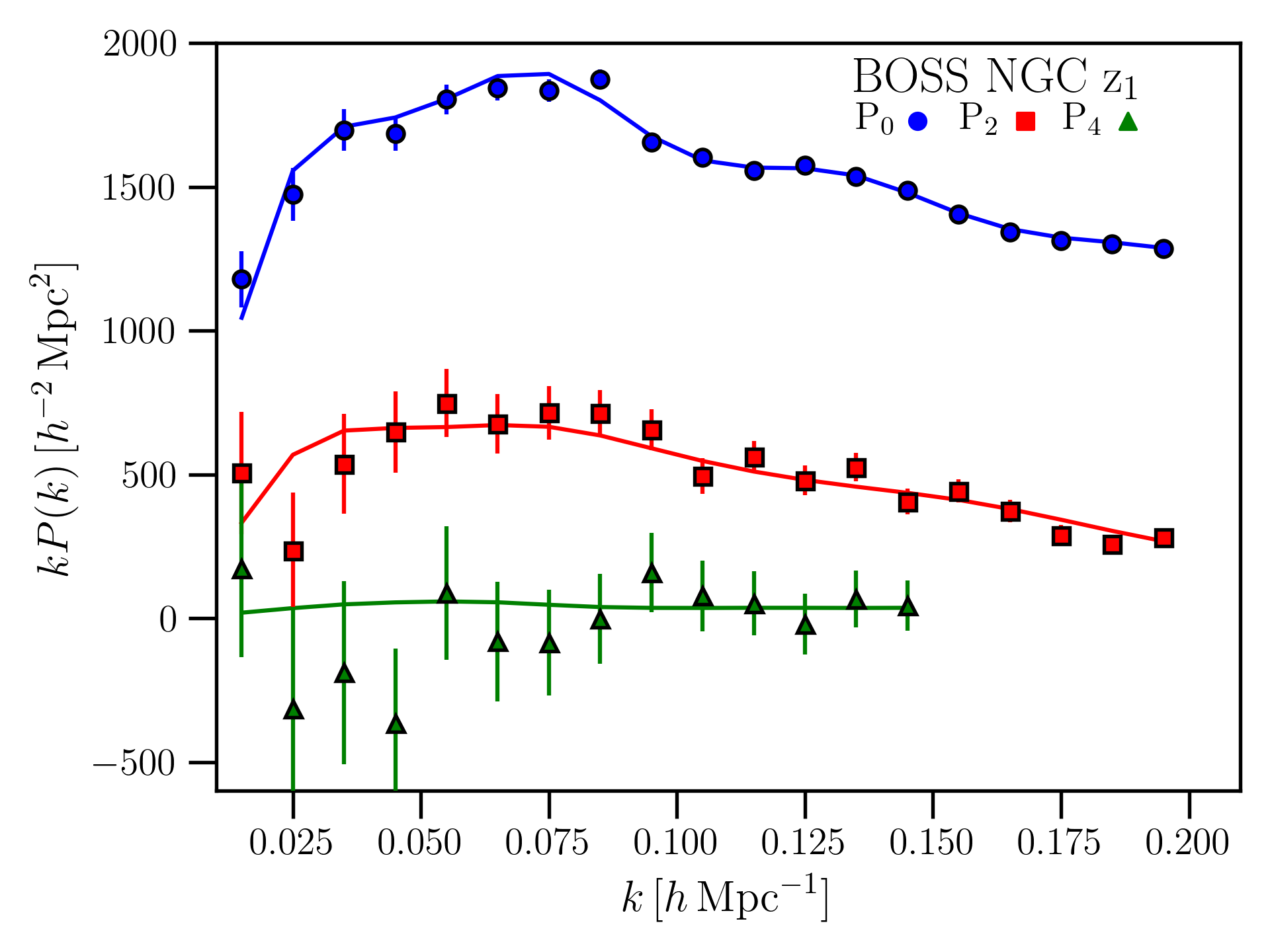}
        \includegraphics[width=5.8cm]{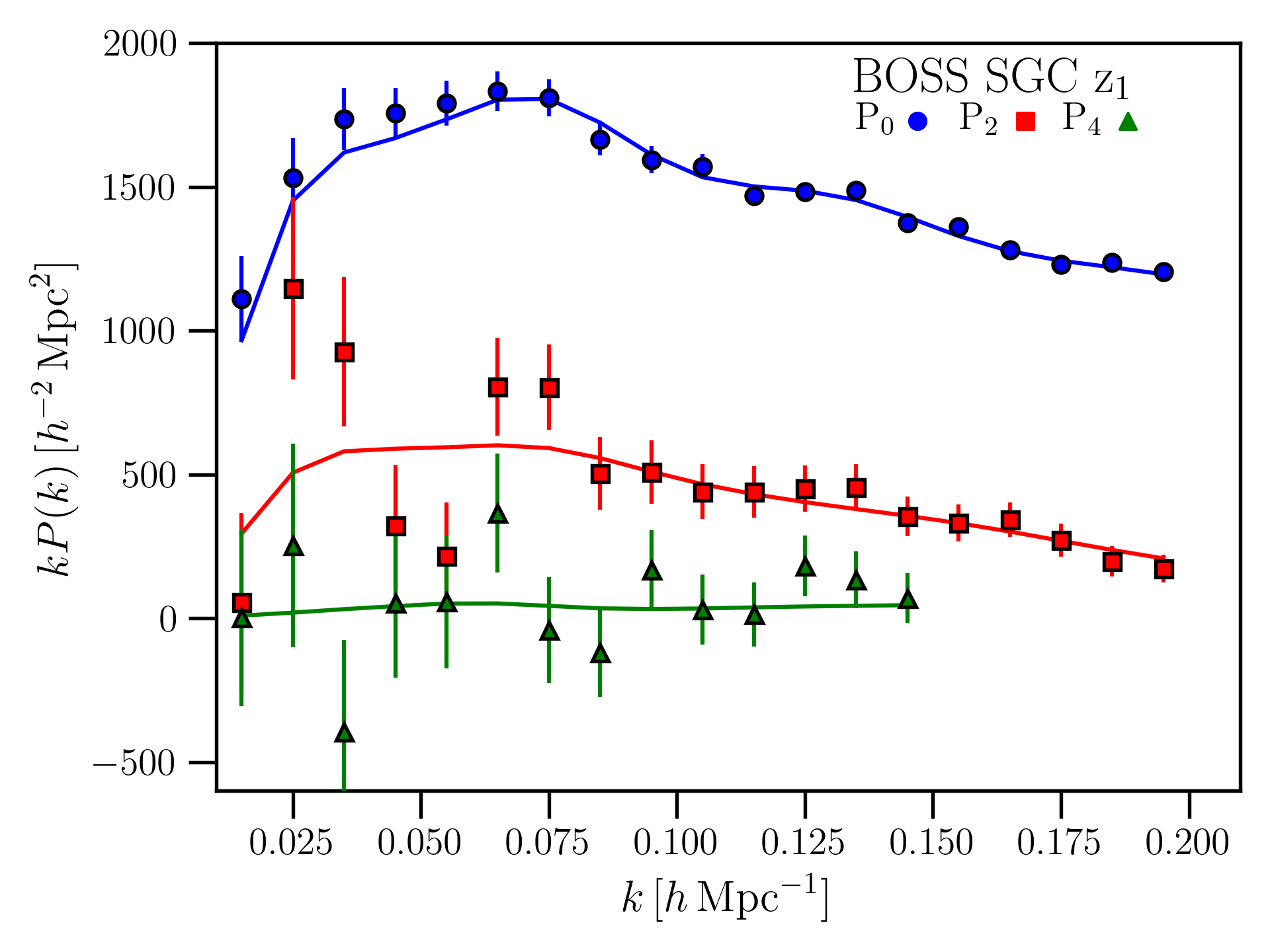}
        \medskip
        \includegraphics[width=5.8cm]{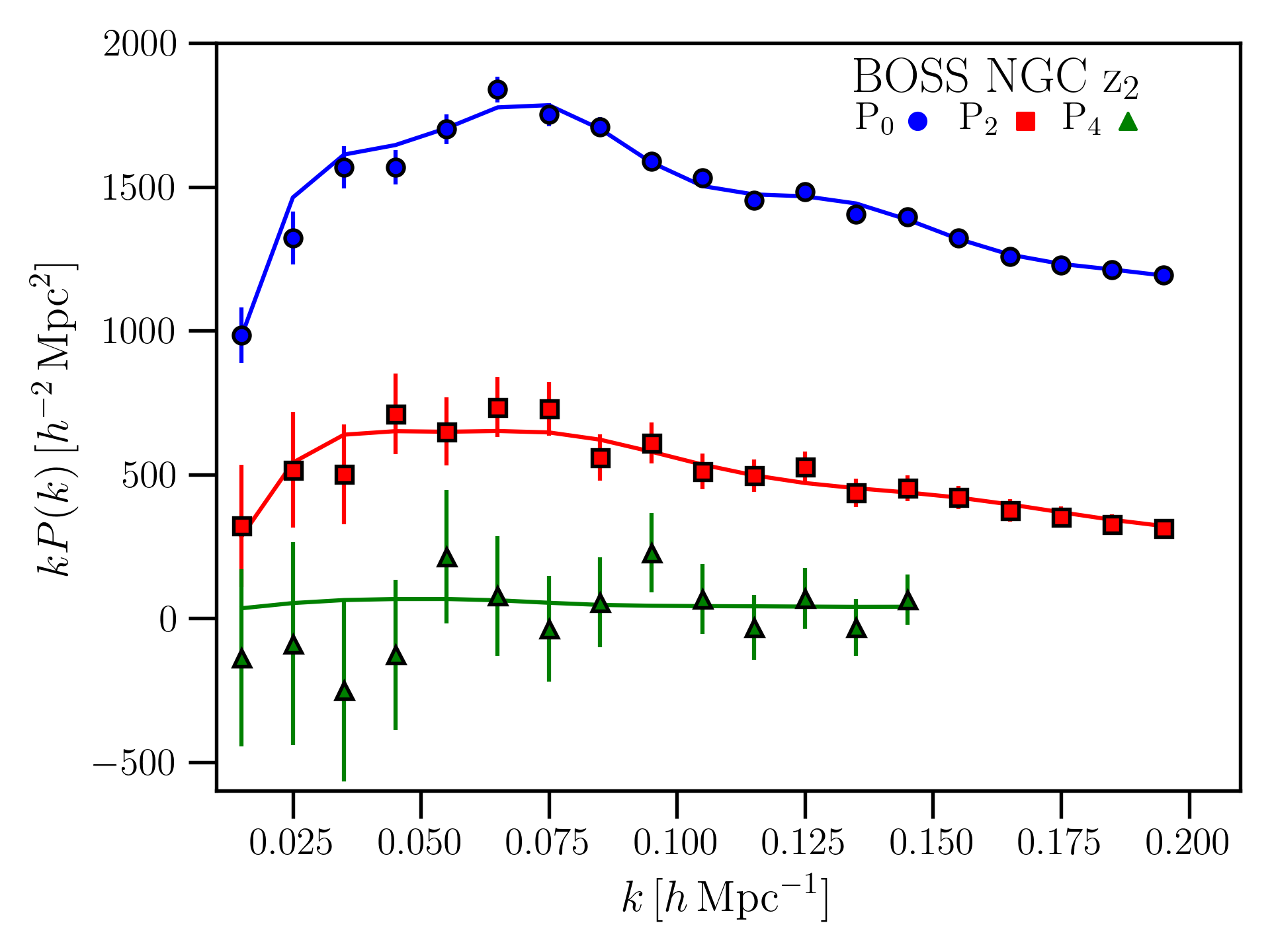}
        \includegraphics[width=5.8cm]{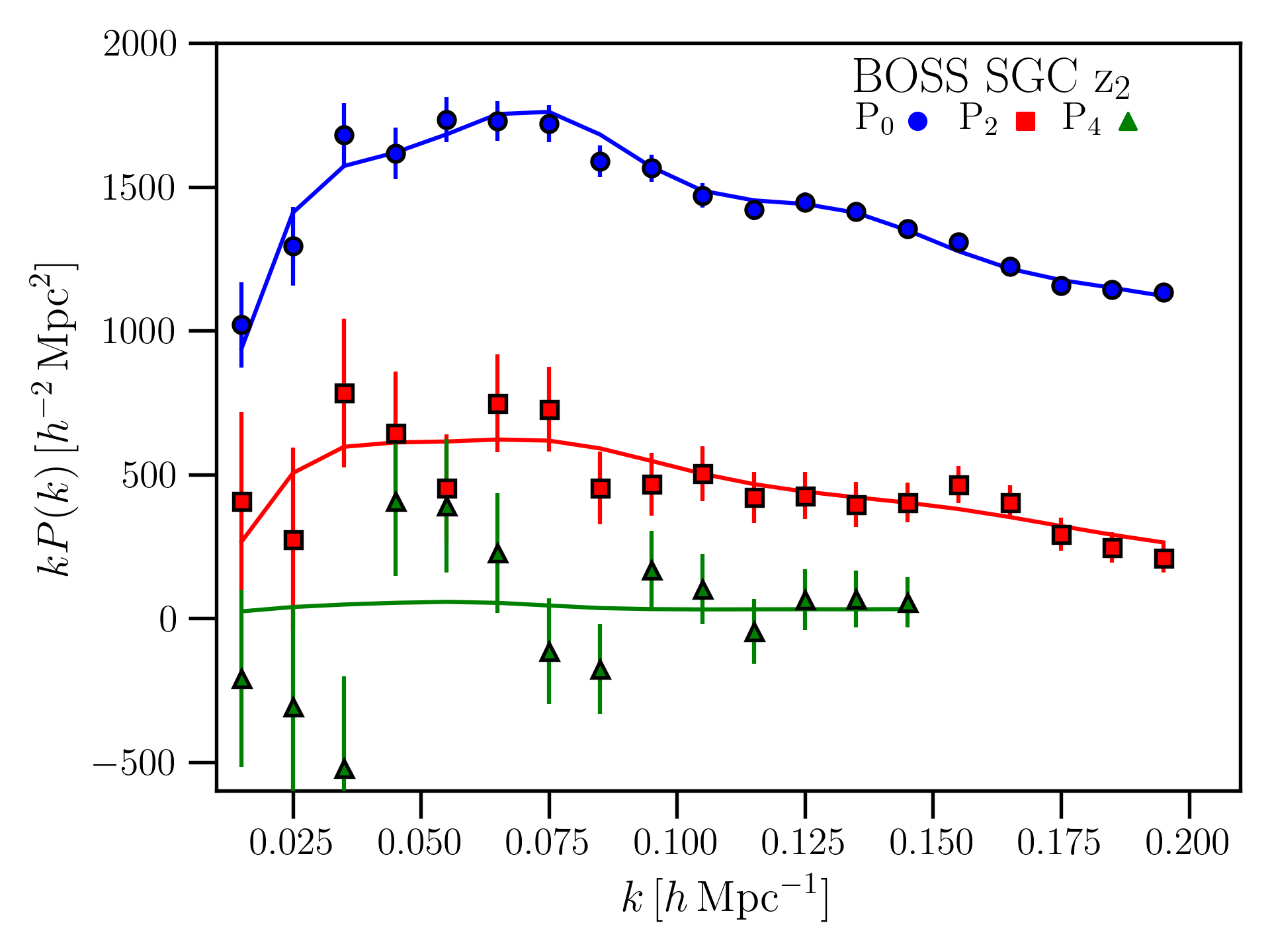}
        \includegraphics[width=5.8cm]{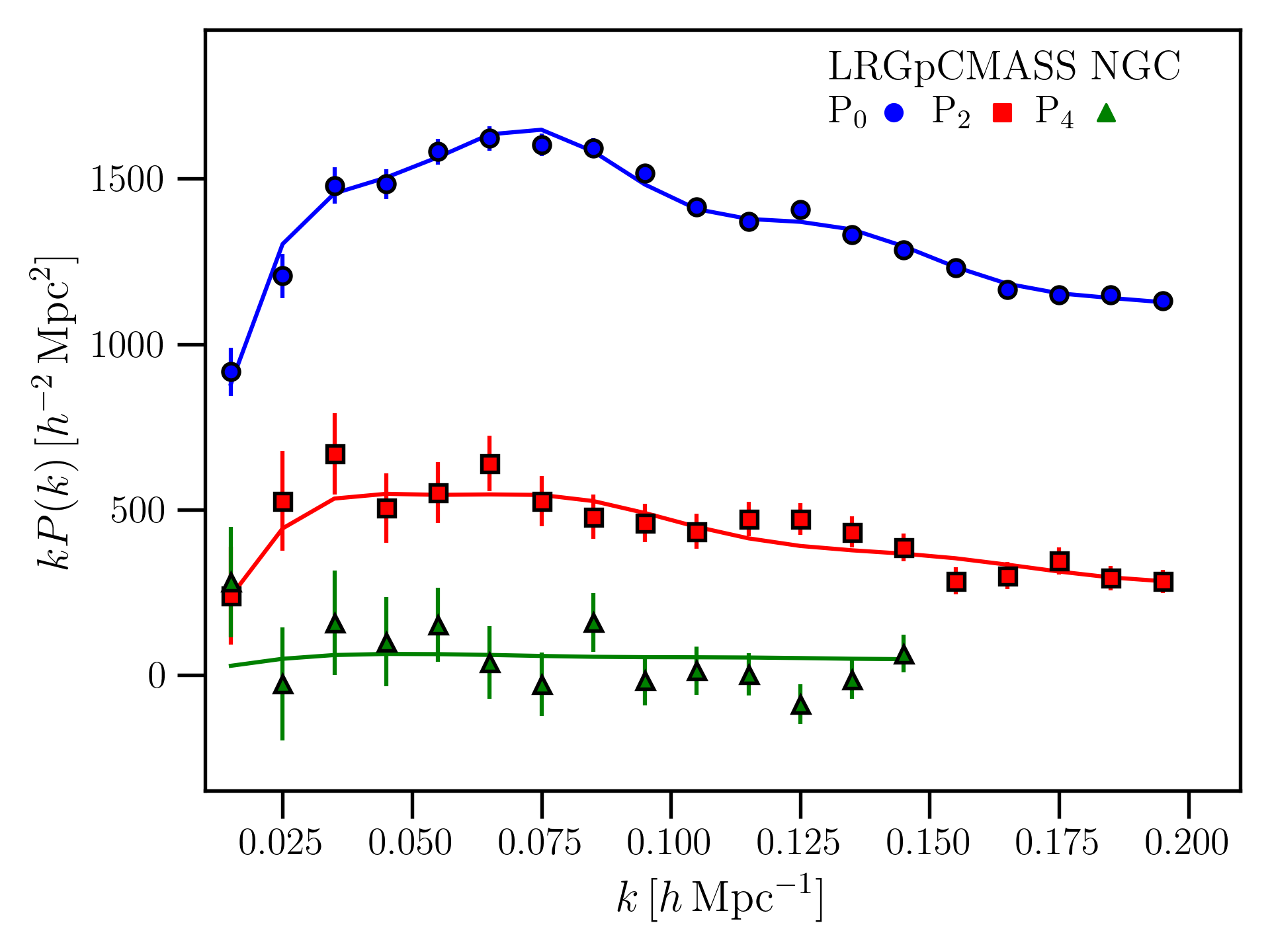}
        \medskip
        \includegraphics[width=5.8cm]{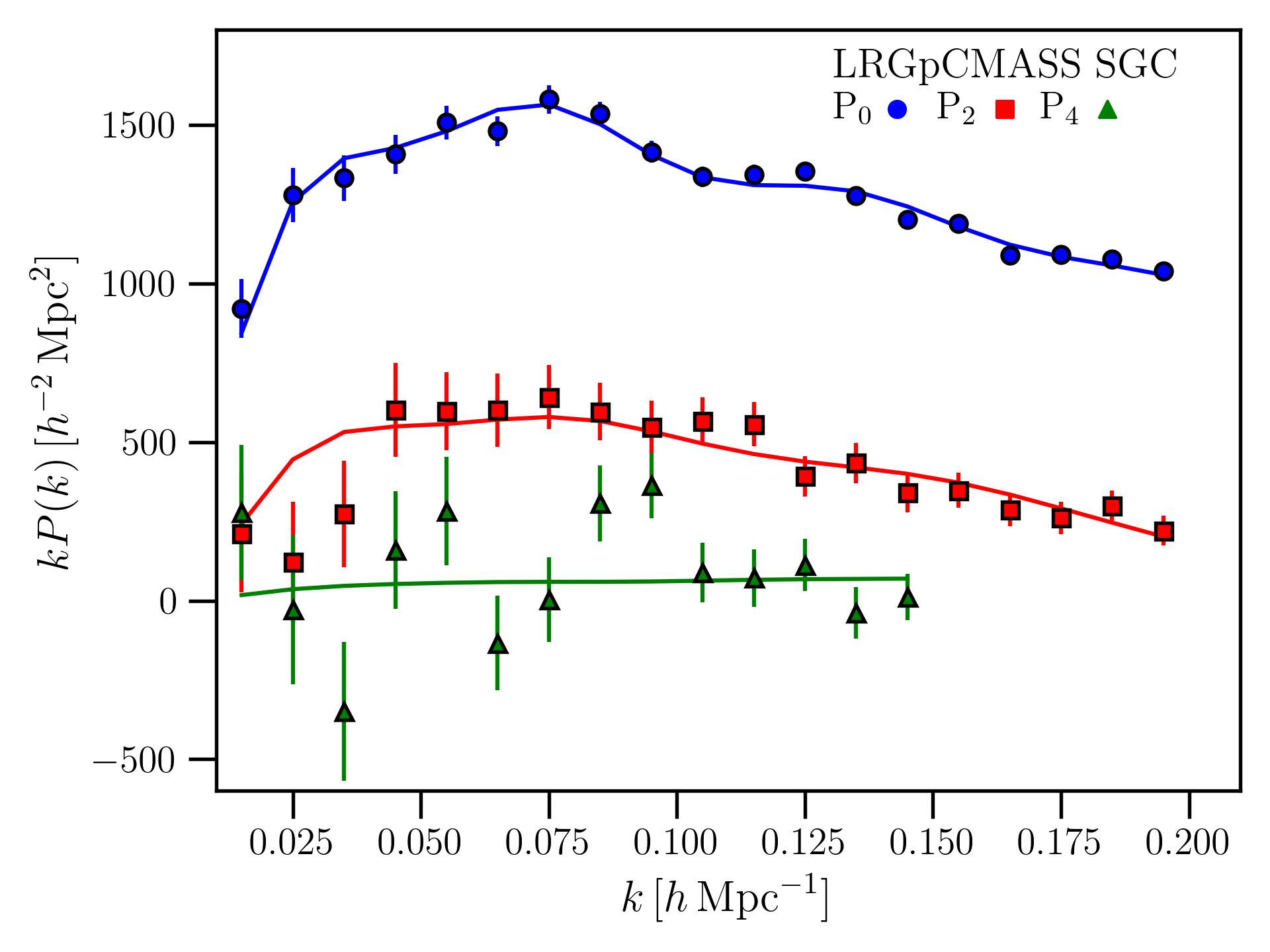}
        \includegraphics[width=5.8cm]{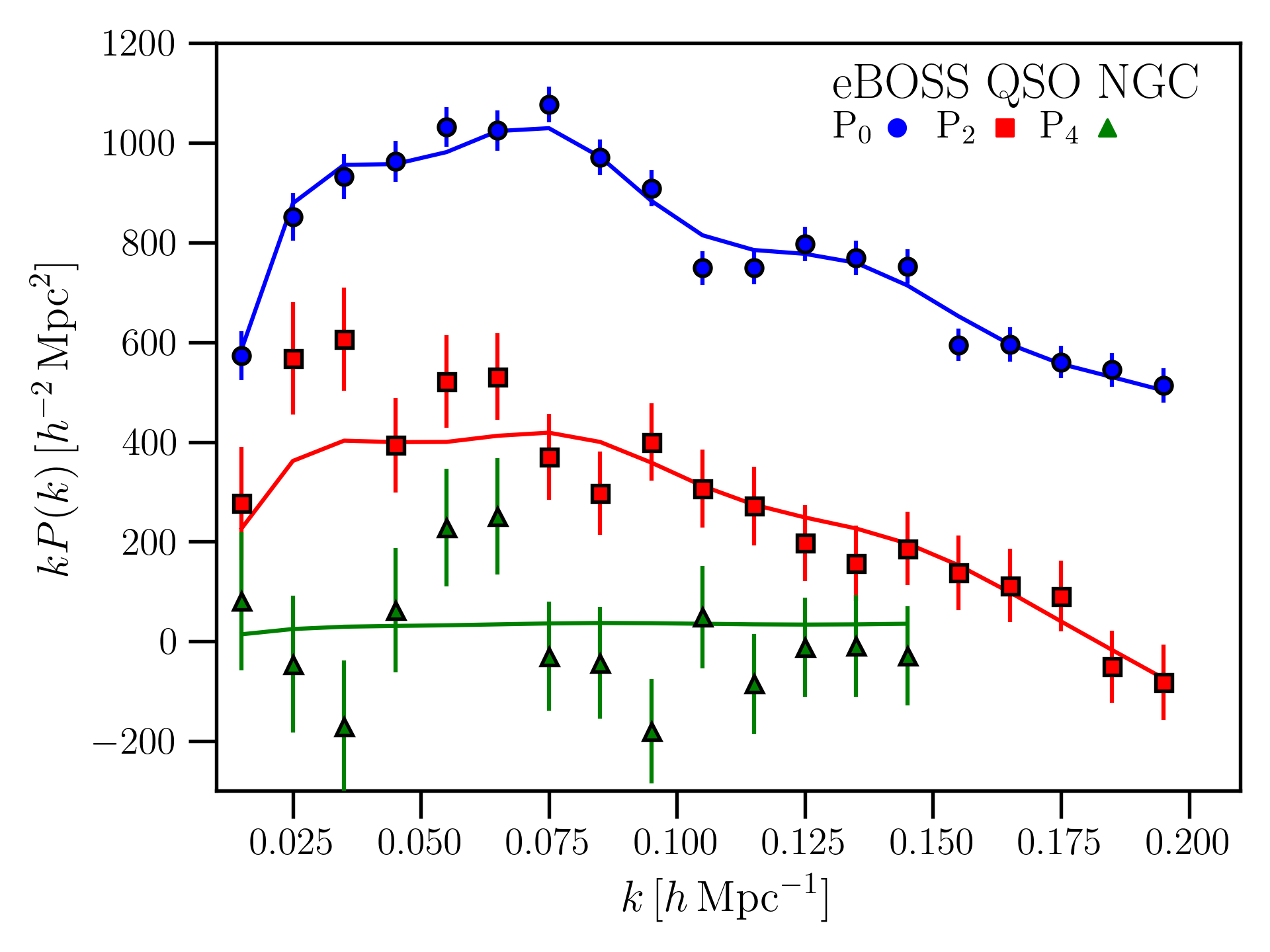}
        \includegraphics[width=5.8cm]{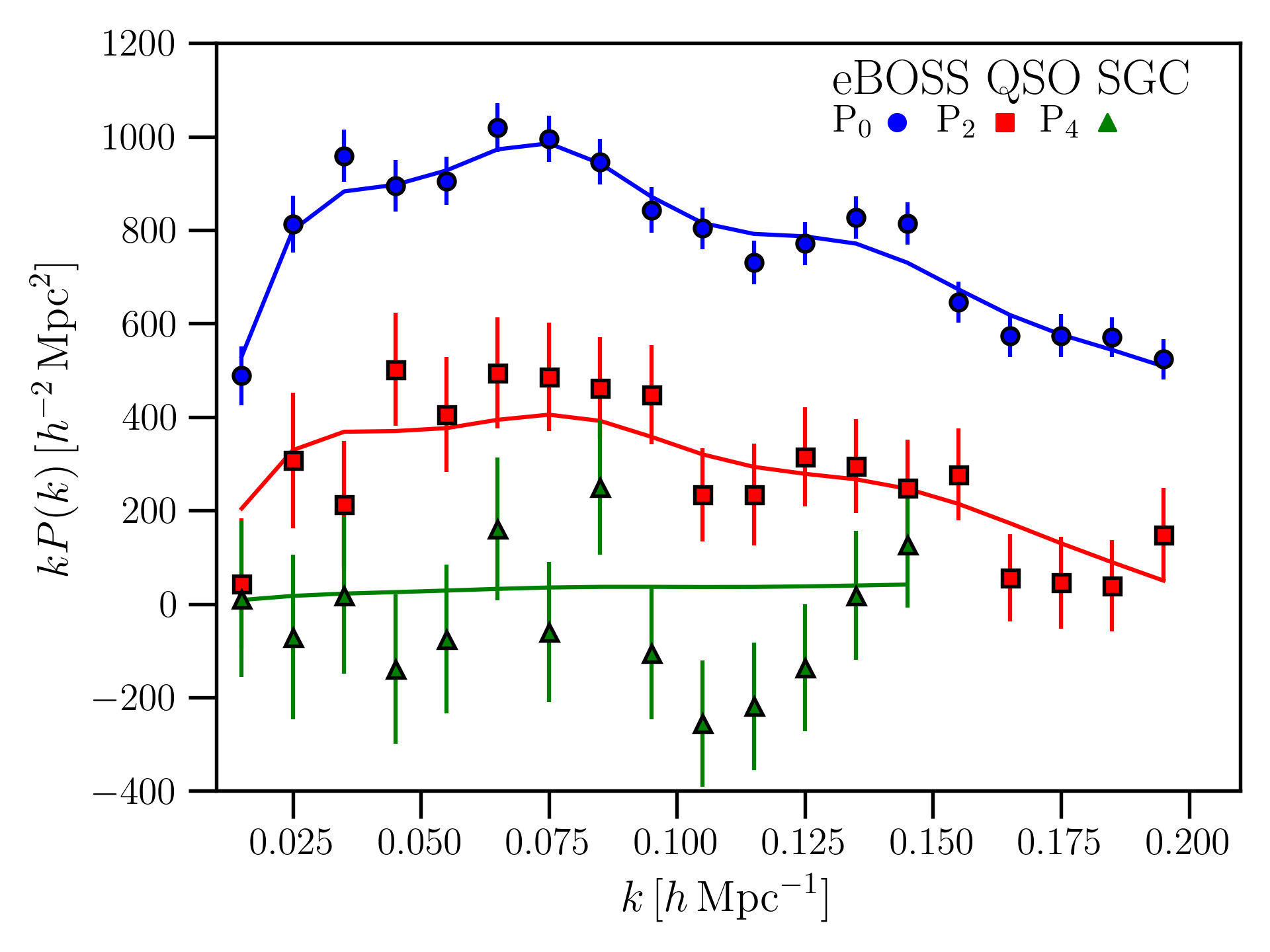}
        \caption[]{Summary of model spectra generated for the 9 survey chunks of our joint 6dFGS+BOSS+eBOSS sample, corresponding to the maximum a posteriori cosmology from our curved run (where $n_s$ is allowed to vary). The model power spectrum for each chunk is derived from our best-fitting ensemble cosmology (given in Table \ref{tab:6dF_BOSS_eBOSS_table}), with independent bias parameters to model the distinct selection/observational effects of each sub-sample. This scheme provides robust fits across the redshift range of our sample, recovering a reduced $\chi^2$ of 1.09, despite the substantial variations in population and selection effects across our suite of measurements.}
        \label{fig:ModelsTrueData}
\end{figure*}

We now turn our attention to employing this validated pipeline to constrain cosmological models using our joint 6dFGS+BOSS+eBOSS sample (following the approach of Section \ref{sec:SamplingStrategy}). We provide the $68\%$ confidence interval across all 4 full-shape only tests in Table \ref{tab:6dF_BOSS_eBOSS_table}, with the full posterior distribution shown in Figure \ref{fig:TrueDataPosterior}. Additionally, we provide the best-fitting model power spectra (across all 9 sub-samples) from our curved, varying $n_s$ run in Figure \ref{fig:ModelsTrueData}.

Combining clustering measurements from 6dFGS, BOSS, and eBOSS, allows us to measure $\Omega_k$ to a precision of $5 \%$ using full-shape information alone, a result which is broadly competitive with other CMB-independent fits in the literature. Relaxing this model and allowing $n_s$ to vary in conjunction with $\Omega_k$ introduces substantial degeneracies in our fits, notably weakening our constraints on $\textrm{ln}(10^{10} A_s)$ and $h$ (parameters which both help modulate the amplitude of our broadband clustering signal). Across both models, our suite of pre-reconstruction full-shape measurements exhibit a $\sim 2 \sigma$ preference for closed cosmologies. This result is notably more significant than the $\sim 1 \sigma$ preference obtained from analysis in \cite{Chudaykin_2020}, where BOSS full-shape measurements were combined with post-reconstruction BAO information. It is unclear whether this difference in marginalised fits to $\Omega_k$ arises due to the impact of post-reconstruction BAO information (which overwhelmingly supports flatness), or from our inclusion of high-redshift clustering information. While beyond the scope of this paper, these results highlight the value of investigating the role of reconstruction in measurements of curvature which utilise BAO information. Finally, it is interesting to note the maximum a posteriori model across our curved fits consistently prefers cosmologies which are $1 \sigma$ closer to flatness than the posterior mean. In addition, we find our goodness of fit (as measured by $\rm{log}(\mathcal{L})$) remains broadly consistent across analyses--- this suggests the inclusion of $\Omega_k$ and $n_s$ in our full-shape models does not substantially improve the quality of our fits. 

\subsubsection{Evidence Ratio}

With the $\textrm{log}(\mathcal{L})$ of our full-shape constraints providing some insight into the relative performance of curved and flat cosmological models, we now turn our attention to developing this model comparison using the evidence ratio. We use nested sampling to compute the Bayesian evidence given by our full-shape measurements when used to constrain flat and curved cosmologies (using a fixed $n_s$). Using our posterior evaluation to refine our nested sampling bounds, we apply the following hard priors for our full-shape model comparison:
\begin{equation}
\begin{split}
    \textrm{ln}(10^{10} A_s) = \{1.5, 3.5\} \hspace{2cm} h = \{0.6, 0.75\} \\
    \Omega_{\textit{cdm}} h^2 = \{0.09, 0.16\} \hspace{1.1cm} \Omega_{b} h^2 = \{0.02, 0.024\} \\
    \Omega_{k} = \{-0.25, 0.15\}  \hspace{3.9cm} 
\end{split}
\end{equation}
While only $\ln(\mathcal{Z})$ is required for this comparison, we additionally quote the  Kullback-Leibler divergence ($\mathcal{D}$) and Bayesian model dimensionality ($d$) for each model. The summary of our results is provided in Table \ref{tab:EvidenceRatio}.

\begin{table}[H]
\centering
\small
\begin{tabular}{lllll}
\hline
\textbf{} & $\textrm{ln}(\mathcal{Z})$ & \textrm{$\mathcal{D}$} & \textrm{$d$} \\ \hline
\textbf{Flat} & $-224.51 \pm 0.24$ &  $31.39 \pm 0.23$ & $22.68 \pm 0.68$ \\
\textbf{Curved} & $-223.87 \pm 0.22$ & $30.80 \pm 0.22$ & $25.35 \pm 0.86$ \\
\hline
\end{tabular}
\caption{Summary of results from nested sampling analysis of our full-shape data in flat and curved models, with derived statistics.}
\label{tab:EvidenceRatio}
\end{table}

\begin{table*}[t]
\fontsize{7}{11}\selectfont
\centering
\renewcommand{\arraystretch}{1.7} 
\setlength{\tabcolsep}{3.0pt}
\begin{tabular}{@{}lllllllll@{}}
\toprule
 & $\rm{ln}(10^{10} A_{\textit{s}})$ & $h$ & $\Omega_{\textit{cdm}} h^2$ & $\Omega_{\textit{m}}$ & $\Omega_k$ & $n_s$ & $2*\rm{log}(\mathcal{L})$ \\ \midrule
 Planck, flat & $3.045^{+0.017}_{-0.016} \ (3.045)$ & $0.679^{+0.007}_{-0.007} \ (0.678)$ & $0.120^{+0.001}_{-0.002} \ (0.120)$ & $0.309^{+0.009}_{-0.009} \ (0.310)$ & - & $0.966^{+0.005}_{-0.004} \ (0.965)$ & $1003.2$ \\
 Planck, curved & $3.030^{+0.018}_{-0.019} \ (3.010)$ & $0.542^{+0.043}_{-0.042} \ (0.529)$ & $0.118^{+0.002}_{-0.002} \ (0.118)$ & $0.470^{+0.076}_{-0.069} \ (0.501)$ & $-0.042^{+0.019}_{-0.021} \ (-0.051)$ & $0.971^{+0.005}_{-0.005} \ (0.972)$ & $994.2$ \\
 \midrule
 Planck + FS, flat & $3.038^{+0.017}_{-0.016} \ (3.041)$ & $0.679^{+0.005}_{-0.005} \ (0.679)$ & $0.120^{+0.001}_{-0.001} \ (0.120)$ & $0.309^{+0.007}_{-0.007} \ (0.308)$ & - & $0.966^{+0.004}_{-0.004} \ (0.968)$ & $1377.7$ \\
 Planck + FS, curved & $ 3.037^{+0.017}_{-0.016} \ (3.035)$ & $0.667^{+0.008}_{-0.009} \ (0.687)$ & $0.118^{+0.001}_{-0.002} \ (0.117)$ & $0.317^{+0.009}_{-0.008} \ (0.317)$ & $-0.0041^{+0.0026}_{-0.0021} \ (-0.0052)$ & $ 0.969^{+0.005}_{-0.005} \ (0.968)$ & $1373.5$\\
 \midrule
\end{tabular}
\caption{$68\%$ confidence interval and best fit provided from full-shape analysis of the Planck TTTEEE lite likelihoods (i.e. with foreground parameters marginalised), both individually and under combination with our full-shape sample via Metropolis-Hastings sampling. Note that during our analysis of the Planck likelihoods, we assume a cosmology with 3 massless Neutrino species (in contrast to the baseline Planck cosmology, which assumes a single massive species of $\Sigma m_{\nu} = 0.06 \ \textrm{eV}$). Our assumption of three massless Neutrino species does not introduce any significant deviation from the standard Planck results, and allows for a more reliable comparison with our full-shape models (which similarly assume a cosmology populated with massless Neutrinos).}
\label{tab:Planck_x_FS_Table}
\end{table*}

We find the full-shape data alone exhibits a very marginal preference for open cosmological models, assigning Bayesian betting odds of $\sim 2:1$ in favour of curvature. This level of support broadly indicates our full-shape measurements are equally compatible with either model, a result supported by the weak improvement in the quality of curved fits as measured by log($\mathcal{L}$).

\subsection{Planck + Full-Shape Analysis}



We now turn our attention to exploring the reliability of cosmological constraints which combine full-shape and Planck information. We make use of the nuisance-marginalised ``Planck-lite'' likelihood for TTTEEE spectra beyond $\ell = 30$, in conjunction with the standard low-$\ell$ TT and EE spectra below this cut-off.  This ``lite'' likelihood marginalises over the standard CMB foreground parameters, encoding these into a single bias parameter $A_{\textrm{Planck}}$. Posterior evaluations which include Planck data follow the default \textsc{MontePython} configuration, which includes a hard prior on $\tau_{\rm{reio}} > 0.004$. All remaining cosmological parameters (including $\Omega_k$) are not defined with explicit priors, allowing our runs to extend over the entire parameter space supported by \textsc{Class}. As previously, our full-shape likelihood includes a Gaussian prior of $100 \times \Omega_b h^2 = 2.235 \pm 0.049$ (informed from \cite{Cooke_2018})--- while this is included in runs which combine full-shape and Planck information, the effect of this prior is marginal in comparison to the $\Omega_b h^2$ fits provided by the Planck likelihood. As outlined in Section \ref{sec:SamplingStrategy} our full-shape cosmological model assumes three massless Neutrino species, in contrast to the single massive Neutrino species (with $\Sigma m_{\nu} = 0.06 \ \textrm{eV}$) used in the baseline Planck model. To ensure our comparison with the full-shape information remains robust, we therefore opt to compute all chains which include Planck information with our three massless neutrino configuration (rather than using the publicly available Planck chains). The full $68\%$ confidence intervals and best fits from fits to the Planck data are provided in Table \ref{tab:Planck_x_FS_Table}.

Reassuringly we find our use of 3 massless neutrino species do not substantially affect our CMB constraints, recovering fits across flat and curved $\Lambda$CDM models which are consistent with the standard Planck analysis. While the posterior distribution from CMB measurements alone are skewed to closed universes at a significance of $\sim 3 \sigma$, including full-shape information yields constraints which are consistent with flatness to within $0.5 \%$. Breaking the geometric degeneracy also brings our measurements of $h$ and $\Omega_m$ into closer alignment with standard cosmological fits.

\begin{figure}
        \includegraphics[width=8.6cm]{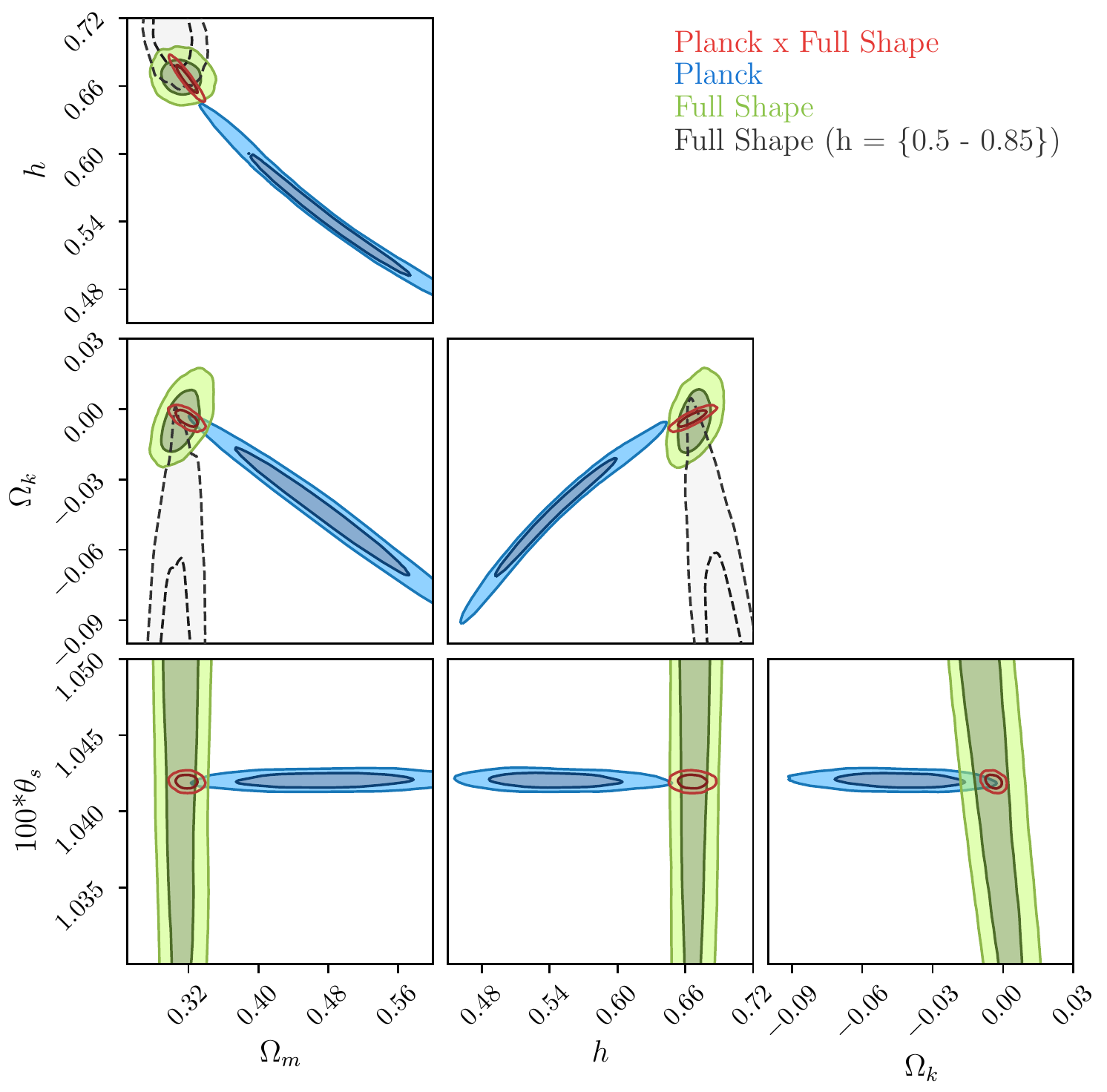}
        \caption[]{Posterior distribution provided from joint full-shape (6dFGS+BOSS+eBOSS) and CMB (Planck 2018) nested sampling fits to curved $\Lambda$CDM models, in contrast with the individual posteriors from both probes. In addition, we include our full-shape posterior evaluated over our wide $h$ prior for reference (Note this is identical to the curved posterior from Figure \ref{fig:TrueDataPosterior}, right). While $\theta_s$ and $\Omega_k$ are directly sampled as part of our evidence calculations, $\Omega_m$ and $h$ are included as derived parameters for clarity. The combination of full-shape and CMB information provides remarkably tight constraints on $\Omega_{k}$, arising due to the intersection of their perpendicular degeneracy directions in $\Omega_{k}$/$\theta_s$ space. While full-shape measurements can accomodate curved cosmologies (as visible through evaluation over our wide $h$ prior), this comes at the expense of a fit to $\theta_s$/$h$ which is strongly excluded by CMB measurements.}
         \label{fig:Planck_plus_FS_posterior}
\end{figure}

While combinations of CMB and full-shape information provides incredibly precise constraints on $\Omega_k$, in order to confidently combine these measurements we must be convinced they are internally consistent. As with our earlier full-shape only analysis, we evaluate this using the Bayesian evidence, this time estimating the congruity of the Planck and full-shape data via the suspicousness statistic. In order to include Planck data in our Nested Sampling runs, we are forced to reduce our prior width on $\Omega_{k}$ to $\{-0.15, 0.1\}$, and perform a change of variables (substituting $100 \times \theta_s$ for $h$), with the full prior range given in Equation \ref{eq:PlanckxFSPrior}. As noted in \cite{Handley2019}, unphysical cosmological combinations at the edges of our prior volume (where the nested sampling run begins) can cause standard cosmology calculations to fail. The full summary of our analysis over this prior range is provided in Table \ref{tab:SuspiciousnessStatistic}, with the posteriors from these nested sampling runs plotted in Figure \ref{fig:Planck_plus_FS_posterior}

 \begin{equation}
\begin{split}
    \ln(10^{10} A_s) = \{1.5, 3.5\} \hspace{0.7cm} 100 \times \theta_s = \{1.025, 1.055\} \\
    \Omega_{\textit{cdm}} h^2 = \{0.09, 0.16\} \hspace{1.2cm} \Omega_{b} h^2 = \{0.02, 0.024\} \\
    n_s = \{0.78 , 1.08\} \hspace{1.65cm} \Omega_{k} = \{-0.15, 0.1\} \\
    \tau_{\rm{reio}} = \{0.004, 0.1\}  \hspace{4cm} \\
    \label{eq:PlanckxFSPrior}
\end{split}
\end{equation}

\begin{table}[H]
\setlength{\tabcolsep}{2.5pt}
\small
\begin{tabular}{lllll}
\hline
\textbf{} & $\textrm{ln}(\mathcal{Z})$ & \textrm{$\mathcal{D}$} & \textrm{$d$} & \textrm{Tension ($\sigma$)} \\ \hline
\textbf{FS} & $-227.41 \pm 0.20$ & $34.26 \pm 0.21$ & $23.45 \pm 0.67$ & - \\
\textbf{Planck} & $-519.97 \pm 0.15$ &  $19.01 \pm 0.14$ & $8.35 \pm 0.24$ & - \\
\midrule
\textbf{Planck + FS} & $-747.41 \pm 0.25$ & $50.72 \pm 0.26$ & $26.28 \pm 0.80$ & $1.76^{+0.14}_{-0.11} \sigma$ \\ \hline
\end{tabular}
\caption{Nested sampling analysis of our full-shape data and Planck information, individually and in combination.}
\label{tab:SuspiciousnessStatistic}
\end{table}

We find evidence for a moderate inconsistency between Planck 2018 and our suite of full-shape measurements, corresponding to a tension probability of $p \sim 0.08 \pm 0.02$, or $1.76^{+0.14}_{-0.11} \sigma$. This inconsistency is identifiable in the joint contours of Figure \ref{fig:Planck_plus_FS_posterior}, where we find joint fits to $\Omega_k$ and $100 \times \theta_s$ are broadly incompatible with our Planck only posterior. While full-shape measurements can accommodate closed cosmologies (as evidenced by our Metropolis-Hastings analysis), this comes at the expense of a $\theta_s$ which is strongly disfavoured by Planck.  Similarly, while Planck alone prefers a low value of $\Omega_{k}$, this comes at the cost of a low $h$ and high $\Omega_{m}$ which is not supported by the full-shape measurements. 

As with any Bayesian statistic, it is essential to consider the choice of prior in our analysis. While our full-shape models are strongly modified by $\Omega_m$ and $h$, the extrapolation of these parameters from $\theta_s$ is highly sensitive to $\Omega_k$. Indeed, despite employing a very broad prior of $100 \times \theta_s = \{1.025 - 1.055\}$ (roughly 100 times the statistical error from Planck), this low-redshift sensitivity enforces an effective $\Omega_k$ prior on our full-shape measurements. Importantly, when our full-shape measurements are combined with Planck information, the resulting joint fits from our nested sampling runs remain wholly consistent with the Metropolis-Hastings analysis provided in Table \ref{tab:Planck_x_FS_Table}. While the traditional Bayes ratio is explicitly prior dependent, our use of the suspiciousness statistic ensures our model comparison remains more robust to this choice of prior.  

Interestingly, the tension measured in our analysis is much less significant than tensions between Planck and other late-universe measurements available in the literature, most notably the ``decisive tension'' between full-shape information (from the BOSS $0.43<z<0.7$ signal) and Planck data measured in \cite{Vagnozzi2020}. While this result is surprising, there are a number of novelties in our analysis which may explain this deviation. Firstly, our full-shape analysis uses the widest range of clustering measurements to date, with redshift measurements from our QSO sample extending out to $z=2.2$. The inclusion of this (relatively) high redshift information from QSO clustering may impact the resulting $\Omega_k / \theta_s$ degeneracy in full-shape information which is so critical to our joint-fits. Indeed, in earlier tests of this work fitting the individual BOSS DR12 redshift bins, we found that the `z3 NGC' sample alone, which is similar to that used in \cite{Vagnozzi2020}, prefers a slightly lower $\Omega_{m}$ than our complete set of full-shape data which could exacerbate the tension somewhat. In addition, we are able to model these clustering statistics to much smaller scales than previous studies, offering many more density modes for analysis. Finally, the impact of our choice of prior (most notably the effect of our $\theta_s$ prior on full-shape measurements in isolation) must be taken into account when considering the significance of this measured tension. We stress that these results should not be interpreted as conclusive evidence against the CMB/late universe tension discussed in the literature. Indeed, until this inconsistency is better understood, joint fits which combine CMB and late universe information (including full-shape clustering measurements) should continue to be treated with caution. Given the novel elements of our analysis however (most notably our inclusion of high-redshift and small-scale clustering information), these results represent an interesting contribution to the continuing curvature tension debate. 

\section{Discussion and Conclusions}

As improvements in measurement precision continue to expose inconsistencies between cosmological datasets, the EFTofLSS provides a powerful new framework to independently revisit measurements across a range of cosmological models. In this paper we investigate the role of full-shape clustering measurements in the emerging curvature tension debate, and explore how common analysis assumptions could affect full-shape curvature constraints. Galaxy clustering statistics frequently assume flatness through the use of reconstruction algorithms, and in the choice of fiducial cosmology used to convert redshifts to distances. We fully disentangle the assumptions of reconstruction from our work by fitting the full-shape of the power spectrum in isolation (i.e. without the inclusion of post-reconstruction BAO information). While our analysis does rely on \textit{some} choice of fiducial cosmology, we quantify the impact of assuming a consistent, non-flat fiducial cosmology across all stages of analysis. Using two extremely curved fiducial cosmologies ($\Omega_{k \textrm{,fid}} = \pm 0.1$), we identify a small scatter with respect to our baseline (i.e. $\Omega_{k, \textrm{fid}} = 0$) validation fits. The resulting offsets are sub-dominant to the expected statistical error on realistic measurements (at a significance of $0.2 \sigma$), and as such are unlikely to bias current full-shape constraints of $\Omega_k$. 

With our pipeline validated against mock catalogues, we constrain flat and curved $\Lambda$CDM cosmologies using a combination of measurements from the 6dFGS, BOSS, and eBOSS samples. Using these full-shape power spectrum measurements alone, we recover robust constraints on flat $\Lambda$CDM cosmologies which are consistent with the Planck results to within $1 \sigma$. When we extend our analysis to include curvature as a free parameter, we recover fits of $\Omega_k$ of $-0.089^{+0.049}_{-0.046}$ (when $n_s$ is fixed) and $\Omega_k = -0.152^{+0.059}_{-0.053}$ (when $n_s$ is allowed to vary). Comparing the relative performance of these models using the Bayesian evidence ratio yields betting odds of 2:1 in favour of curvature, indicating our full-shape data remains broadly compatible with both cosmological models.

When our suite of full-shape measurements are combined with Planck CMB information to jointly constrain curved $\Lambda$CDM models, we recover results which are consistent with flatness to within $0.5\%$ ($\Omega_k = -0.0041^{+0.0026}_{-0.0021}$). Using the suspiciousness statistic we identify a moderate tension between CMB and full-shape measurements in curved $\Lambda$CDM models, with a tension probability of $p \sim 0.08 \pm 0.02$, or $1.76^{+0.14}_{-0.11} \sigma$. Like any Bayesian measurement, this should be understood as the tension recovered given our chosen model parameterisation and prior--- as such, our results should not be understood as broad evidence against the strong CMB/late universe tension recorded in the literature. However, given the novel elements of our analysis (most notably, our use of very high-$z$ and small-scale clustering information), we argue this is an interesting result in the emerging curvature tension debate. It will thus be interesting to explore further with upcoming data from surveys such as DESI \citep{DESI2016}, which our methodology and pipeline enables us to do.

\section*{Acknowledgements}
The authors would like to thank Will Handley and David Parkinson for very useful discussions and correspondence. This research was supported by the Australian Government through the Australian Research Council’s Laureate Fellowship funding scheme (project FL180100168). AG is the recipient of an Australian Government Research Training Program (RTP) Scholarship.

\section*{Data Availability}
The \textsc{PyBird} code used in this analysis is available \href{https://github.com/CullanHowlett/pybird/tree/desi}{here}, and the \textsc{MontePython} likelihoods built on this model are available \href{https://github.com/AaronGlanville/montepython_public/tree/3.4/montepython/likelihoods/pybird_FullShape_Pk}{here}. The Nseries mock catalogues used in this work, alongside the LRGpCMASS samples are available \href{https://www.ub.edu/bispectrum/page12.html}{here}. The remaining BOSS and eBOSS measurements used in this analysis are available from the following \href{https://sas.sdss.org/}{url}, with the 6dFGS catalogues available \href{http://www-wfau.roe.ac.uk/6dFGS/}{here}. Any additional data products will be made available upon reasonable request to the corresponding author.



\bibliographystyle{mnras}
\bibliography{bibliography} 






\bsp	
\label{lastpage}
\end{document}